\documentclass[aps,prl,reprint,superscriptaddress,longbibliography]{revtex4-2}
\PassOptionsToPackage{dvipsnames,svgnames,table}{xcolor}
\usepackage{amsmath,amssymb,amsfonts,bm,bbm}
\usepackage{hyperref}

\usepackage[utf8]{inputenc} 

\usepackage{graphicx}
\usepackage{dcolumn}
\usepackage{bm}
\usepackage{enumerate}
\usepackage{orcidlink}

\usepackage{braket}
\usepackage{amsfonts}
\usepackage{amsmath}
\usepackage{amssymb}

\usepackage{color,soul}
\usepackage{wasysym}
\usepackage{mathrsfs}
\usepackage{times}

\usepackage[dvipsnames,svgnames,table]{xcolor}
\usepackage{hyperref}
\usepackage{fancyhdr}
\usepackage{float}
\usepackage[english]{babel}
\let\babelselectlanguage\selectlanguage
\renewcommand{\selectlanguage}[1]{%
  \def\requestedlanguage{#1}%
  \def\enlanguage{en}%
  \def\ENlanguage{EN}%
  \def\Enlanguage{En}%
  \ifx\requestedlanguage\enlanguage
    \babelselectlanguage{english}%
  \else\ifx\requestedlanguage\ENlanguage
    \babelselectlanguage{english}%
  \else\ifx\requestedlanguage\Enlanguage
    \babelselectlanguage{english}%
  \else
    \babelselectlanguage{#1}%
  \fi\fi\fi}
\usepackage[caption=false]{subfig}
\usepackage{braket}

\usepackage{graphicx}
\hypersetup{
  breaklinks = {true},
  citecolor = {blue},
  colorlinks = {true},
  linkcolor = {red},
}
\usepackage{physics}
\usepackage{mathtools}
\usepackage{braket} 
\usepackage{upgreek}

\usepackage[normalem]{ulem} 

\newcommand{\Tof}{T_{\mathrm{of}}}
\newcommand{\Db}{\Delta_b}

\hypersetup{
pdfstartview={FitH},
colorlinks=true,    
linkcolor=NavyBlue, 
citecolor=Maroon,   
filecolor=NavyBlue, 
urlcolor=NavyBlue   
}

\usepackage{array}
\usepackage{booktabs}

\newcommand{\bea}{\begin{eqnarray}}
\newcommand{\eea}{\end{eqnarray}}

\begin{document}

\title{Precision-Induced Irreversibility in non-Hermitian systems}

\author{Luis E. F. Foa Torres}
\affiliation{Departamento de F\'{\i}sica, Facultad de Ciencias F\'{\i}sicas y Matem\'aticas, Universidad de Chile, Santiago, Chile}
\email{luis.foatorres@uchile.cl} 
\author{Giorgos Pappas}
\affiliation{Laboratoire d'Acoustique de l'Universit\'e du Mans (LAUM), UMR 6613, Institut d'Acoustique - Graduate School (IA-GS), CNRS, Le Mans Universit\'e, Av. Olivier Messiaen, 72085 Le Mans, France}
\author{Vassos Achilleos}
\affiliation{Laboratoire d'Acoustique de l'Universit\'e du Mans (LAUM), UMR 6613, Institut d'Acoustique - Graduate School (IA-GS), CNRS, Le Mans Universit\'e, Av. Olivier Messiaen, 72085 Le Mans, France}
\author{Diego Bautista Avilés}
\affiliation{Departamento de F\'{\i}sica, Facultad de Ciencias F\'{\i}sicas y Matem\'aticas, Universidad de Chile, Santiago, Chile}
\date{\today}

\begin{abstract}
Non-Hermitian evolution is mathematically invertible, yet finite dynamic range imposes a sharp operational limit on reversibility. We identify Precision-Induced Irreversibility (PIR): amplification, mode mixing (as warranted by non-normality), and a finite resolution floor---whether set by detector noise, environmental fluctuations, or numerical precision---conspire to produce a quantitative predictability horizon $T_{\mathrm{of}}$, beyond which distinct states collapse onto identical representations. Within the effective non-Hermitian description, the mechanism requires neither environmental decoherence nor nonlinear dynamics; remove any ingredient and reversibility can be restored. Echo-fidelity tests confirm this transition across arbitrary-precision arithmetic and hardware, revealing where formal invertibility and physical reversibility diverge.

\end{abstract}

\maketitle

\textit{Introduction.--} How does irreversibility emerge from reversible equations? From molecular gases to quantum devices, time-symmetric microscopic laws coexist with robust macroscopic arrows of time. Two mechanisms dominate this realm: environmental decoherence destroys quantum coherence through entanglement with additional degrees of freedom~\cite{zurek_decoherence_2003}, while deterministic chaos exponentially amplifies small uncertainties until trajectories become unpredictable~\cite{elskens_instability_1986, gutzwiller_m_c_chaos_1990}. Quantum systems whose classical counterparts are chaotic inherit this sensitivity, as revealed by the Loschmidt echo~\cite{jalabert_environment-independent_2001}. In both cases, irreversibility is tied to either environmental entanglement or nonlinear dynamics.

\begin{figure}[pth]
\centering
\includegraphics[width=0.95\columnwidth]{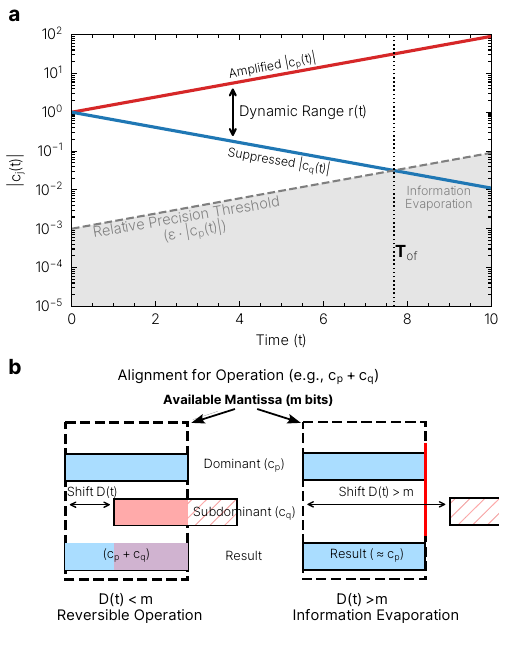}
\caption{\textbf{The dynamic-range crisis and predictability horizon.}
(a) Under non-Hermitian evolution, the amplified mode $|c_p(t)|$ (red) grows while suppressed mode $|c_q(t)|$ (blue) decays exponentially, driving dynamic range ratio $r(t) = |c_p|/|c_q|$ across orders of magnitude. The shaded area (precision shadow) marks the relative precision threshold $\varepsilon \cdot |c_p(t)|$ below which the subdominant component becomes numerically unresolvable. At the  overflow time $T_{\text{of}}$, $|c_q|$ falls into the precision shadow, forcing a many-to-one mapping and information evaporates.
(b) Many-to-one mapping mechanism. Left: Before $T_{\text{of}}$, the available precision bits $m$ suffice to represent both components ($D(t) < m$). Right: After $T_{\text{of}}$, the dynamic range overflows the precision capacity ($D(t) > m$), and distinct initial states collapse to identical representations.}
\label{fig1}
\end{figure}

What about strictly linear dynamics? Two facts limit the possible answers: the propagator $e^{-i{\cal H}t/\hbar}$ is mathematically invertible (even when ${\cal H}$ is non-Hermitian!), while physical precision is finite. Because unitarity preserves all scales equally, Hermitian evolution harmonizes these two aspects. However, non-Hermitian evolution does not—even though the propagator remains perfectly invertible. Decades of research have explored PT-symmetry \cite{bender_real_1998}, exceptional points~\cite{ashida_non-hermitian_2020,bergholtz_exceptional_2021,li_exceptional_2023}, their cycling dynamics~\cite{berry_slow_2011,uzdin_observability_2011,nye_universal_2023,nye_adiabatic_2024,kumar_general_2025,wang_chiral_2026}, phenomena such as chiral state conversion~\cite{doppler_dynamically_2016,xu_topological_2016,hassan_chiral_2017,foa_torres_non-hermitian_2025} and the non-Hermitian skin effect~\cite{martinez_alvarez_non-hermitian_2018,yao_edge_2018,lin_topological_2023}, and non-normality~\cite{trefethen_spectra_2005}. Yet while non-normality~\cite{trefethen_spectra_2005,feng_numerical_2025} has been identified as a source of severe numerical instabilities in non-Hermitian computations, the physical implications of this breakdown remain unexplored.

Here we identify \emph{Precision-Induced Irreversibility} (PIR)---a route to the arrow of time within non-Hermitian dynamics. Like decoherence and chaos-based irreversibility, PIR is an emergent phenomenon at the level of description where experiments operate. Unlike them, it arises from the interplay of three ingredients in strictly linear evolution: amplification, mode mixing (as warranted by non-normality), and finite dynamic range. These form an essential trinity; remove any one and reversibility can be restored.
Non-Hermitian systems with gain and loss~\cite{rotter_non-hermitian_2009, moiseyev_non-hermitian_2011}, now realized across photonics~\cite{ruter_observation_2010} and other wave platforms, provide the natural setting. In these classical wave platforms the non-Hermitian equation is the fundamental equation of motion---there is no reservoir to reverse---so the predictability horizon is intrinsic; in open quantum systems the same ${\cal H}$ arises as a reduced description and the horizon appears at the accessible level. Exponential amplification drives amplitudes apart, creating a dynamic-range crisis (Fig.~\ref{fig1}a), much as a camera sensor struggles to capture a scene against a very bright background. It turns out amplification alone is not enough: non-normality forces the mixed-scale additions of Fig.~\ref{fig1}b, without which each component could be tracked separately. When the subdominant component sinks below the precision threshold, distinct states collapse onto identical representations and information evaporates. Unlike Landauer erasure~\cite{landauer_irreversibility_1961, plenio_physics_2001}, where bits are deliberately discarded, PIR describes information that becomes unresolvable because the non-unitary dynamics overwhelms the representation capacity. It also differs from Prigogine's precision-limited irreversibility in (classically) chaotic systems~\cite{prigogine_end_1997}: there nonlinear dynamics demands infinite precision to track trajectories, whereas PIR requires no chaos and operates within strictly linear evolution. The minimal two-level setting places PIR beyond the reach of chaos-based irreversibility, which requires nonlinear dynamics and continuous phase space. More broadly, Del Santo and Gisin have argued that infinite-precision is not physically realizable---the Bekenstein bound forbids encoding unlimited information in finite volume~\cite{delsanto_physics_2019, gisin_mathematical_2020}. PIR provides a quantitative dynamical formula for the consequences of this finiteness in linear non-Hermitian systems.

We derive a predictability horizon $T_{\mathrm{of}}$, scaling linearly with the precision bits $m$ and inversely with the amplification rate $\Delta b$ (the argument in $\ln(\beta)$ is the numerical base, typically $\beta=2$). Beyond identifying this irreversibility mechanism, the relation can also be inverted: measuring $T_{\mathrm{of}}$ in an echo experiment directly yields the \emph{effective number of bits} encoded in a given photonic, electronic, or mechanical platform. Reversibility persists for $t < T_{\mathrm{of}}$ and collapses sharply afterward, when the finite-precision dynamics and analytical predictions generally part ways. We verify this prediction using two independent observables: the Loschmidt-echo (fidelity)~\cite{peres_stability_1984, goussev_loschmidt_2016} and work-echo ratio—both detecting the same $T_{\mathrm{of}}$. 
The transition remains consistent across arbitrary-precision calculations and standard floating-point hardware, showing that PIR represents a true physical limitation rather than a numerical artifact. Beyond the echo protocol, the same precision-limited timescale also governs eigenstate selection in slow-driven non-Hermitian systems~\cite{pappas_universal_2026}, confirming that PIR manifests in forward-only evolution as well.

\paragraph*{Quantifying the dynamic-range crisis.}

To illustrate the mechanism behind PIR, it is useful to start from a
minimal two–mode gain/loss problem.
Consider the time evolution of a state under the Schr\"odinger equation
with a non-Hermitian \(2\times 2\) Hamiltonian of the form
\begin{equation}
    i\hbar \,\partial_t |\psi(t)\rangle ={\cal H}\,|\psi(t)\rangle,
    \qquad {\cal H} = K - i\Gamma,
\end{equation}
where \(K^\dagger = K\) and \(\Gamma^\dagger = \Gamma\) introduce gain or
loss.  In a basis where \({\cal H}\) has diagonal entries with imaginary
parts \(\hbar b_j(t)\), the two components obey
\(|c_j(t)| \propto \exp\!\bigl[-\int_0^t b_j(\tau)d\tau\bigr]\).  The
dynamic-range ratio between an amplified component \(p\) and a suppressed
component \(q\) is therefore
\begin{equation}
    r(t) \equiv
    \frac{|c_p(t)|}{|c_q(t)|}
    = r(0)\,
      \exp\!\left[\int_0^t \Delta b(\tau)\,d\tau\right],
    \qquad
    \Delta b \equiv b_q - b_p ,
    \label{eq:r-of-t}
\end{equation}
which grows exponentially with the integrated
\(\Delta b\). This component-wise form is a heuristic motivation valid in a basis where $\mathcal{H}$ is effectively diagonal; the basis-independent condition-number formulation below provides the rigorous, general framework.

The quantity that matters for PIR is not overall magnitude but the \emph{relative} dynamic range between components:
\begin{equation}
    D(t) = \frac{1}{\ln 2}\int_0^t \Delta b(\tau)\,d\tau,
    \label{eq:D-of-t}
\end{equation}
which counts accumulated dynamic-range growth in bits. PIR onset occurs when $D(t)$ exceeds the available precision $m$; global scale changes alone never cause irreversibility as long as they are tracked.

Let the representation capacity be characterized by a minimal resolvable
\emph{relative} scale
\begin{equation}
    \varepsilon = \beta^{-m},
\end{equation}
where \(\beta\) is the numerical base (typically \(\beta=2\)) and \(m\) is the
available precision. In numerical simulations \(m\) is literally
the precision bit-width; in experiments it represents an effective dynamic
range \(m_{\mathrm{eff}} = \mathrm{DR(dB)}/(20\log_{10}2) \approx \mathrm{DR(dB)}/6\),
set by detector SNR, amplifier saturation, or quantization depth rather than
floating-point format (see the Supplemental Material~\cite{supp_material} for the mapping).
As long as \(r(t)\ll \varepsilon^{-1}\), both components are
faithfully resolved.  Once the dynamic-range ratio exceeds the inverse
precision, \(r(t)\gtrsim \varepsilon^{-1}\), the subdominant component is
driven below the relative threshold and underflows: any operation that
adds the two components (Fig.~\ref{fig1}b) only
``sees'' the larger one, and previously distinguishable states collapse
to the same representation.

In a diagonal (normal) system, each channel evolves independently and can be inverted separately, so dynamic-range growth alone does not cause irreversibility. However, it defines the time scale at which the resolution floor $\varepsilon$ is exhausted. From $D(T) \simeq \log_2(1/\varepsilon)$ with constant $\Delta b$:
\begin{equation}
    T_{\mathrm{DR}} \simeq \frac{\ln(1/\varepsilon)}{\Delta b}
    = \frac{m\ln \beta}{\Delta b}\,,
    \label{eq:Tof-DR}
\end{equation}
where the last equality uses the precision floor $\varepsilon = \beta^{-m}$. Although $\varepsilon$ is set by the arithmetic precision in simulations, the same expression applies to any resolution floor (e.g.\ detector noise or quantization depth).
True PIR emerges when evolution necessarily mixes amplified and suppressed components---as non-normality guarantees---so that sub-precision losses cannot be undone channel by channel.

\paragraph*{Basis-independent formulation via the propagator condition number.}

To make this estimate independent of basis choice and initial state,
and to connect directly with the echo protocol, we recast the same
threshold in terms of the full propagator \(U(t)\) generated by the
non-Hermitian evolution.  We define the condition number~\cite{trefethen_spectra_2005}
\begin{equation}
    \kappa\!\left(U(t)\right)
    \equiv \|U(t)\|\,\|U(t)^{-1}\|
    = \frac{\sigma_{\max}(U(t))}{\sigma_{\min}(U(t))},
\end{equation}
where \(\sigma_{\max/\min}\) are the maximum and minimum singular values in any
unitarily invariant norm.  For Hermitian evolution,
\(\kappa\equiv 1\); for gain/loss dynamics in the broken phase,
\(\kappa\bigl(U(t)\bigr)\) grows approximately exponentially,
\(\kappa\!\left(U(t)\right)\simeq C\,\exp(\Delta b\, t)\), with the same
growth rate \(\Delta b\) as in Eq.~\eqref{eq:r-of-t} and a geometric
prefactor \(C\propto\kappa(V)^{2}\) set by eigenvector
non-orthogonality (exact form in the Supplementary Information).

Any initial perturbation \(\delta\psi_0\) with relative size
\(\|\delta\psi_0\|/\|\psi_0\|\lesssim\varepsilon\) is amplified according
to
\begin{equation}
  \frac{\|\delta\psi(t)\|}{\|\psi(t)\|}
  \le \kappa\!\left(U(t)\right)\,
      \frac{\|\delta\psi_0\|}{\|\psi_0\|}
  \lesssim \kappa\!\left(U(t)\right)\,\varepsilon .
\end{equation}
Finite precision therefore remains operationally harmless as long as
\(\kappa(U(t))\,\varepsilon\ll 1\).
We define the overflow (or predictability) horizon \(T_{\mathrm{of}}\)
by the condition
\begin{equation}
    \kappa\!\left(U(T_{\mathrm{of}})\right)\,\varepsilon \sim c ,
    \label{eq:kappa-threshold}
\end{equation}
where \(c=\mathcal{O}(1)\) specifies the precise ``knee'' criterion
(e.g.\ a fixed drop of the Loschmidt echo).
It is convenient to re-express this in terms of a ``condition-number
register''
\begin{equation}
    D_\kappa(t) \equiv \log_{\beta} \kappa\!\left(U(t)\right)
    = \frac{\ln\kappa(U(t))}{\ln \beta},
\end{equation}
which counts how many base-\(\beta\) digits of dynamic range the propagator
has accumulated.  Eq.~\eqref{eq:kappa-threshold} is then
equivalent to
\begin{equation}
    D_\kappa\bigl(T_{\mathrm{of}}\bigr)
    \simeq m + \log_{\beta} c .
\end{equation}
With \(\kappa(U(t))\simeq C\,\exp(\Delta b t)\) we have
\(D_\kappa(t)\simeq (\Delta b/\ln \beta)\,t + \log_\beta C\), and thus
\begin{equation}
    T_{\mathrm{of}}
    \simeq T_{\mathrm{DR}} - \frac{\ln C}{\Delta b}
    = \frac{m\ln \beta - \ln C}{\Delta b}\,,
    \qquad C\propto\kappa(V)^{2}.
    \label{eq:Tof-final}
\end{equation}
The scaling \(T_{\mathrm{of}}\propto m\) is
\emph{initial-state independent}: only the propagator's singular-value ratio and the precision floor enter the threshold condition. The correction \(-\ln C/\Delta b\) is a constant time shift, independent of \(m\), that advances the overflow time: eigenvector non-orthogonality effectively amplifies the resolution floor from \(\varepsilon\) to \(C\,\varepsilon\), so that fewer bits of dynamic range are available before overflow.

More generally, $T_{\mathrm{DR}} = \ln(1/\varepsilon)/\Delta b$ applies to any resolution floor, including componentwise precision errors ($\varepsilon = \beta^{-m}$) and global perturbations (environmental or detector noise); non-normality is required for the former while the $\ln C/\Delta b$ shift persists for both (see Supp. Information).

\begin{figure}
\centering
\includegraphics[width=\columnwidth]{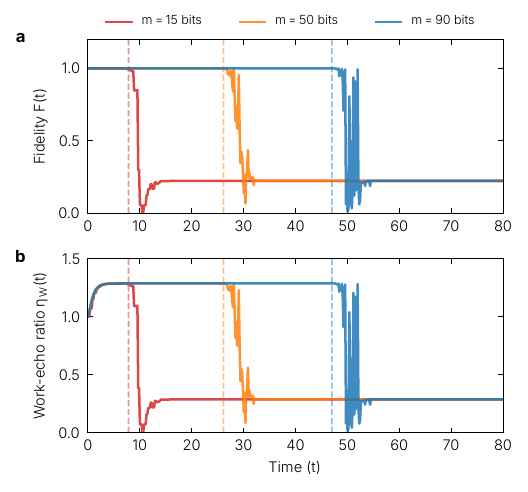}
\caption{\textbf{Sharp-knee signatures of precision-induced reversibility breakdown.} Initial state $|\psi_0\rangle \propto (1,r_0)^T$ with $r_0 = 10^{-2}$, normalized (predominantly on the gain site). (a) Loschmidt-echo fidelity $F(t)$ for precision values $m \in \{15, 50, 90\}$ bits using mpmath with time step $\Delta t = 0.4$. Sharp transition at $T_{\text{of}}$ separates reversible ($F \approx 1$) from irreversible ($F \ll 1$) regimes. Vertical dashed lines mark predicted $T_{\text{of}} = m\ln(\beta)/\Delta b$ with $\Delta b = 1.327$ (time in units of $1/g$). (b) Work-echo ratio $\eta_W(t)$ for the same precision values, showing three regimes: $m$-independent reversible plateau $\eta_W^{\text{rev}} \approx 1.2$ for $t < T_{\text{of}}$, sharp knee at $T_{\text{of}}$, and precision- and state-independent saturation $\eta_W^{\infty} \approx 0.3$ for $t \gg T_{\text{of}}$.}
\label{fig:signatures}
\end{figure}

\begin{figure}
\centering
\includegraphics[width=0.98\columnwidth]{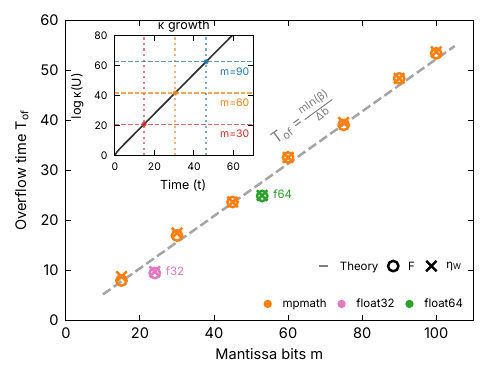}
\caption{\textbf{Universal scaling of overflow time with precision.}
  Initial state $|\psi_0\rangle \propto (1,r_0)^T$ with $r_0 = 10^{-2}$, normalized; $T_{\mathrm{of}}$ is initial-state independent.
  Main panel: Overflow time $T_{\text{of}}$ versus precision bits $m$, measured via Loschmidt fidelity $F$ (circles) and work-echo
  ratio $\eta_W$ (crosses) using onset detection (1\% deviation from reversible plateau). Data from arbitrary-precision stepped
  evolution (mpmath, orange) and native hardware arithmetic (float32,
  pink; float64, green) follow the theoretical prediction $T_{\text{of}} \propto m\ln(\beta)/\Delta b$ (dashed line) with $\Delta b = 2\sqrt{\gamma^2 - g^2} = 1.327$ for $\gamma=1.2$, $g=1.0$.
  Inset: $\ln\kappa(U(t))$ grows exponentially at rate $\Delta b$, crossing precision thresholds $\ln\kappa_{\text{th}} = m\ln(\beta)$ at the predicted $T_{\text{of}}$ for $m \in \{30, 60, 90\}$ bits.}
\label{fig:quantification}
\end{figure}

\textit{The echo protocol.--}
How do we make the abstract dynamic-range crisis visible in actual
observables?  To this end we employ a precision
stress test that turns the loss of relative information into measurable
echo signatures.

The protocol unfolds in two stages.  First, we prepare an initial state
$|\psi_0\rangle$ and evolve it under a non-Hermitian Hamiltonian
$\mathcal H$ for a time $\tau$,
\begin{equation}
    |\psi(\tau)\rangle = U(\tau)\,|\psi_0\rangle,
    \qquad
    U(\tau) = e^{-i\mathcal H \tau/\hbar}.
\end{equation}
In this forward leg the gain/loss bias $\Delta b$ steadily stretches the
relative register and increases the condition number
$\kappa\!\left(U(\tau)\right)$.  Second, we attempt a return by applying
the \emph{inverse} evolution generated by $-\mathcal H$,
\begin{equation}
    U_{\mathrm{b}}(\tau) = e^{+i\mathcal H \tau/\hbar}
    = U(\tau)^{-1}
\end{equation}
in exact arithmetic, and ask whether the initial state can be
recovered.

We use two diagnostics to probe reversibility.  The
\emph{Loschmidt-echo or fidelity} tests whether the dynamics can retrace
its steps:
\begin{equation}
    F(\tau) =
    \frac{\bigl|\langle\psi_0|
           U_{\mathrm{b}}(\tau)\,U(\tau)
           |\psi_0\rangle\bigr|^2}
         {\langle\psi_0|\psi_0\rangle\,
          \langle\psi_{\mathrm{rec}}(\tau)|\psi_{\mathrm{rec}}(\tau)\rangle},
\end{equation}
with $|\psi_{\mathrm{rec}}(\tau)\rangle \equiv U_{\mathrm{b}}(\tau)\,U(\tau)|\psi_0\rangle$. For perfectly reversible evolution in exact arithmetic one has $F=1$;
representation-induced information loss drives $F$ away from this ideal
value.

As a complementary, thermodynamic-style readout we consider a
\emph{work-echo ratio}
\begin{equation}
    \eta_W(\tau) \equiv
    \frac{W_{\mathrm{rec}}(\tau)}{W_{\mathrm{out}}(\tau)}.
\end{equation}
Here $W_{\mathrm{out}}(\tau)$ and $W_{\mathrm{rec}}(\tau)$ are defined from expectation values of a
chosen ``readout'' Hamiltonian $H_0$ (shifted by the ground-state energy to ensure $W \geq 0$; see Supplementary Information for details).
When information remains intact, $\eta_W(\tau)$ stays near a baseline
value $\eta_W^{\mathrm{rev}}$ set by the initial state and readout
choice.  Once information has evaporated, the
work-echo ratio collapses.

Figure~\ref{fig:signatures} shows the fidelity [panel~(a)] and the
work-echo ratio [panel~(b)] across three decades in precision,
$m \in \{15, 50, 90\}$ bits.  For early times, $\tau < T_{\mathrm{of}}$,
the echo is essentially perfect: $F(\tau)\approx 1$ and
$\eta_W(\tau) \approx \eta_W^{\mathrm{rev}}$, with no visible dependence
on $m$. The plateau value $\eta_W^{\mathrm{rev}}$ depends on the choice of initial state $|\psi_0\rangle$.  At later times both
diagnostics exhibit a sharp drop; the long-time behavior is
$m$-independent and, crucially, also \emph{initial-state independent}: different preparations all converge to the same asymptotic regime, determined by eigenmode structure alone. This universality is the signature of information loss—the system has forgotten which state it started from.

Importantly, both diagnostics spotlight the \emph{same} overflow time
$T_{\mathrm{of}}$. The near-coincidence of the full curves, beyond sharing the onset, reflects a common origin: both diagnostics are governed by $\kappa(U(\tau))$. The threshold and transition width ($\sim 1/\Delta b$) are universal as a worst-case bound, while plateau and asymptotic values depend on the observable (initial state, readout Hamiltonian). This occurs when subdominant components have fallen below the resolution floor, so that
distinct initial states are funneled into identical finite-precision
representations.  Once two preparations have produced the same discrete
trajectory up to $\tau$ at a fixed precision $m$, no further evolution
within that representation can distinguish them: the information about their differences has evaporated.

Figure~\ref{fig:quantification} provides quantitative validation. The inset shows $\ln\kappa(U(\tau))$ growing exponentially at rate $\Delta b$, crossing precision thresholds at the predicted $T_{\mathrm{of}}$ values. The main panel confirms the linear scaling $T_{\mathrm{of}} = m\ln(\beta)/\Delta b$ across arbitrary-precision arithmetic, explicit quantization models, and native floating-point hardware (\texttt{float32}, \texttt{float64}). Both fidelity and work-echo diagnostics yield consistent $T_{\mathrm{of}}$ values, establishing PIR as a genuine physical threshold independent of numerical implementation.

\textit{Scope and relation to decoherence.--} The non-Hermitian Hamiltonian $\mathcal{H}$ may be the fundamental dynamical equation of a classical wave platform (photonic waveguides, acoustic resonators, electrical circuits with gain-loss elements), or an effective description of an open quantum system obtained by projecting out environmental degrees of freedom. PIR operates identically in both cases: it depends on the propagator's condition number and the resolution floor $\varepsilon$, not on the Hamiltonian's provenance. In quantum open systems, PIR coexists with decoherence as an independent channel at the reduced-system level: decoherence produces gradual fidelity decay set by the bath coupling, while PIR produces a sharp threshold at $T_\mathrm{of}$ set by $m$. In classical wave platforms this question does not arise: the non-Hermitian equation is fundamental, there is no reservoir and no reduction, and PIR is the genuine and only irreversibility. In open quantum systems $\mathcal{H}_\mathrm{eff}$ is a reduced description and the $\kappa(U_\mathrm{eff})$ horizon coexists with reservoir inaccessibility---the Hermitian parent has $\kappa(U_\mathrm{full})=1$ and a perfect full-system echo. The no-jump trajectory shows only that this horizon is carried by the deterministic conditional generator: $\kappa(U_\mathrm{eff})$ grows whether or not jumps occur, so PIR is not an artifact of omitting the jump/noise channel (confirmed by re-introducing continuous noise explicitly; see Supplementary Information)~\cite{pan_non-hermitian_2020, deng_stability_2021}. Reservoir inaccessibility, however, is the premise of every effective open-system description, decoherence included---decoherence is itself reversed once the environment is made accessible, as in spin-echo and quantum-eraser protocols---so what distinguishes PIR is not the eventual fate of the information but the qualitatively different signature of its loss at the accessible level: a sharp, $m$-controlled threshold rather than a smooth, bath-controlled decay.

\textit{Experimental proposal.--}
The signatures of precision-induced irreversibility could be observed in a coupled-waveguide dimer: two evanescently coupled waveguides, one with optical gain and the other with matched loss, realizing the PT-symmetric Hamiltonian where propagation distance plays the role of time. Such PT-symmetric dimers have been demonstrated in integrated photonics~\cite{ruter_observation_2010}. The Loschmidt echo requires backward evolution under $-{\cal H}$, which for this system is achieved exactly by the operation $i\sigma_y$, swapping the waveguides while introducing a relative $\pi$ phase shift.

In the proposed linear cavity geometry, a partial mirror at the input allows light injection and signal extraction, while a specially designed end reflector implements $i\sigma_y$ through a waveguide crossing~\cite{chen_low-loss_2006} combined with a $\pi$ phase element~\cite{wang_integrated_2018}. Forward and backward propagation traverse the same physical structure, eliminating fabrication mismatches. The key experimental signature is the scaling $T_{\mathrm{of}} \propto 1/\Delta b$, testable by varying the gain-loss contrast while monitoring echo fidelity. An alternative electrical circuit platform with programmable bit truncation would enable direct verification of the complementary scaling $T_{\mathrm{of}} \propto m$. Details are provided in the Supplementary Information.

\textit{Final remarks.}-- Precision-Induced Irreversibility is an emergent irreversibility within non-Hermitian dynamics, operationally distinct from environmental decoherence and deterministic chaos in temporal signature, control parameters, and structural requirements. Amplification, mode mixing (as warranted by non-normality), and finite dynamic range form an essential trinity; eliminating any one allows reversibility to be restored. The predictability horizon $T_{\mathrm{of}} = T_{\mathrm{DR}} - \ln C/\Delta b$, with $T_{\mathrm{DR}} = \ln(1/\varepsilon)/\Delta b$, marks a sharp boundary beyond which distinguishable configurations become computationally indistinguishable. Because $T_{\mathrm{DR}}$ applies to any resolution floor (hardware noise, environmental fluctuations, or finite precision), the mechanism is universal, with the geometric correction $\ln C/\Delta b$ from eigenvector non-orthogonality advancing the overflow uniformly. In this sense, $T_{\mathrm{of}}$ supplies, within the class of amplifying linear systems studied here, a covariant criterion of the kind Wheeler called for in 1978~\cite{wheeler_frontiers_1979}: a quantity depending only on intrinsic properties of the dynamics ($\Delta b$) and
the substrate's available resolution ($\varepsilon$)---an apparatus parameter, itself measurable from the echo---not on observer-chosen thresholds for what counts as an irreversible amplification.

Finite dynamic range is a physical constraint~\cite{delsanto_physics_2019}, not merely a technical limitation: since any noise floor bounds the number of physically determined digits, the predictability horizon is inescapable for non-normal dynamics within any finite-precision realization. On a physical platform the resolution floor is set by shot noise, amplifier saturation, or quantization---not a software precision that can be raised at will---so the horizon $T_{\mathrm{of}}$ is a property of the apparatus, not a bookkeeping choice. PIR reveals where formal invertibility and operational reversibility diverge: the governing equations permit time-reversal, yet no physical implementation can execute it. The echo protocol turns this limitation into a resource: by monitoring where reversibility fails, one reads out the effective number of bits a given physical platform can faithfully represent; calibrating the echo timescale recovers this effective bit-depth to $\sim 1$ bit, with sensitivity $\mathrm{d}m_{\mathrm{eff}}/\mathrm{d}T_{\mathrm{of}}=\Delta b/\ln\beta$ (Supplementary Information). As Borges wrote~\cite{borges_ficciones_1944}, ``To think is to forget a difference.'' Finite dynamic range enforces this forgetting, and what physics cannot distinguish, it cannot reverse.

\textit{Data Availability.--}
The data and code that support the findings of this article are openly available~\cite{foatorres_zenodo_2026}.

\textit{Acknowledgments.--}
We thank Igor Gornyi, Alexander Mirlin and Ihor Poboiko for useful discussions, Hernán Calvo, Cecilia Cormick, Gonzalo Usaj, Lucas Fernández-Alcázar, Alba Ramos, Felipe Barra and Rodrigo Soto for useful comments. L.E.F.F.T. acknowledges financial support by ANID FONDECYT (Chile) through grant 1250751, The Abdus Salam International Centre for Theoretical Physics and the Simons Foundation. D.B.A. acknowledges the financial support of ANID/Subdirección de Capital Humano through Beca Doctorado Nacional Chile/21250325. V.A. and G.P. are supported by the EU H2020 ERC StG ``NASA'' Grant Agreement No. 101077954.

\textit{Author Contributions.--} This work emerged from a collaboration on non-Hermitian dynamics between L.E.F.F.T. and V.A., later joined by G.P. and D.B.A. L.E.F.F.T. conceived the present study, developed the theoretical framework, performed numerical simulations, and wrote the manuscript. D.B.A. performed numerical simulations and contributed key discussions during the initial development of the ideas. V.A. and G.P. contributed to discussions. All authors revised the manuscript.

\onecolumngrid
\clearpage

\begin{center}
\textbf{\large Supplementary Information for:\\[2pt]
``Precision-Induced Irreversibility in non-Hermitian systems''}
\end{center}

\renewcommand{\thefigure}{S\arabic{figure}}
\renewcommand{\thetable}{S\arabic{table}}
\renewcommand{\theequation}{S\arabic{equation}}
\renewcommand{\thesection}{\Alph{section}}
\renewcommand{\thesubsection}{\Alph{section}.\arabic{subsection}}
\setcounter{figure}{0}
\setcounter{table}{0}
\setcounter{equation}{0}
\setcounter{section}{0}

\setcounter{secnumdepth}{3}

\section{Why PIR is Fundamentally Non-Hermitian}
\label{SI:hermitian}

Precision-Induced Irreversibility requires three ingredients: \emph{amplification}, mode mixing (which is warranted by \emph{non-normality}), and \emph{finite precision}. Two condition numbers characterize the first two: $\kappa(U)$, the propagator condition number, measures amplification (how much the evolution stretches some directions relative to others); $\kappa(V)$, the eigenvector condition number, measures non-normality (how non-orthogonal the eigenvectors are). This section explains why \emph{both} are necessary, and why non-normality is a generic factor that separates unavoidably irreversible dynamics from those where reversibility can be maintained.

\subsection{Two Condition Numbers: Amplification and Non-Normality}

For a Hamiltonian $\mathcal{H}$ with eigenvector matrix $V$ (satisfying $\mathcal{H} = V\Lambda V^{-1}$), we define:

\paragraph{Propagator condition number $\kappa(U)$.} For the propagator $U(t) = e^{-i\mathcal{H}t}$,
\begin{equation}
    \kappa(U(t)) = \frac{\sigma_{\max}(U(t))}{\sigma_{\min}(U(t))},
\end{equation}
where $\sigma_{\max/\min}$ are the largest and smallest singular values. This measures how much $U(t)$ amplifies some directions relative to others. For unitary (Hermitian) evolution, $\kappa(U) = 1$ always. For non-Hermitian systems with gain/loss, $\kappa(U)$ can grow exponentially.

\paragraph{Eigenvector condition number $\kappa(V)$.}
\begin{equation}
    \kappa(V) = \|V\| \, \|V^{-1}\|,
\end{equation}
which measures how non-orthogonal the eigenvectors are (see Ref.~\cite{S:trefethen_spectra_2005} for a comprehensive treatment). For normal operators (including all Hermitian operators), eigenvectors are orthogonal and $\kappa(V) = 1$. For non-normal operators, eigenvectors can be oblique, giving $\kappa(V) > 1$.

\subsection{Hermitian Systems: Doubly Protected}

In Hermitian systems, the time-evolution operator $U(t) = e^{-i\mathcal{H}t}$ is \textbf{unitary}, satisfying $U^\dagger U = I$. This provides two protections:

\begin{enumerate}
    \item \textbf{No amplification:} All singular values equal unity, so $\kappa(U) = 1$ for all times. Errors do not grow:
    \begin{equation}
        \|U(t)\,\delta\psi\| = \|\delta\psi\|.
    \end{equation}

    \item \textbf{Orthogonal eigenvectors:} Hermitian operators are normal, so $\kappa(V) = 1$. There is no error leakage between modes.
\end{enumerate}

Together, these ensure that finite precision affects only \textbf{accuracy} (gradual drift from the true trajectory) but not \textbf{stability} (errors remain bounded). Time reversal always succeeds:
\begin{equation}
    U(-t)\,U(t) = I \quad \text{(exact, in any precision)}.
\end{equation}

\subsection{Non-Hermitian Systems: Amplification Creates Vulnerability}

When $\mathcal{H} \neq \mathcal{H}^\dagger$, the propagator $U(t) = e^{-i\mathcal{H}t}$ is no longer unitary. For systems with gain and loss, singular values grow and shrink exponentially:
\begin{equation}
    \sigma_{\max}(U) \sim e^{(\Db/2) t}, \qquad \sigma_{\min}(U) \sim e^{-(\Db/2) t},
\end{equation}
where $\Db$ is the eigenvalue gap (the difference between the imaginary parts of the eigenvalues, i.e., $\Db = 2\sqrt{\gamma^2 - g^2}$ for the PT-symmetric dimer, with $\gamma$ the gain/loss strength and $g$ the coupling). The propagator condition number grows as
\begin{equation}
    \kappa(U(t)) \sim e^{\Db t}.
\end{equation}

This exponential growth of $\kappa(U)$ creates vulnerability to precision errors. A small initial error $\varepsilon$ can be amplified to $\kappa(U) \cdot \varepsilon$. When this product exceeds order unity, errors overwhelm the signal. This defines the \textbf{dynamic-range timescale}:
\begin{equation}
    T_{\mathrm{DR}} = \frac{\ln(1/\varepsilon)}{\Db}\,,
\end{equation}
where $\varepsilon$ is the resolution floor. For numerical simulations with $m$ mantissa bits in base $\beta$ (typically $\beta = 2$), $\varepsilon = \beta^{-m}$ gives $T_{\mathrm{DR}} = m\ln\beta/\Db$. For experiments, $\varepsilon$ is set by the noise floor. The actual \textbf{overflow time} is shifted earlier by eigenvector non-orthogonality: $\Tof = T_{\mathrm{DR}} - \ln C/\Db$ with $C \propto \kappa(V)^2$ [see Sec.~\ref{sec:exact_solution} for the exact expression].

\textbf{However, amplification alone is not sufficient for PIR.} As we show below, if the eigenvectors remain orthogonal ($\kappa(V) = 1$), reversibility is preserved even when $\kappa(U)$ grows exponentially.

\subsection{The Mechanism}

The crucial role of mode mixing (enabled generically by non-normality) becomes clear when we consider how errors propagate through the eigenvector transformation.

\paragraph{Normal systems ($\kappa(V) = 1$).} When eigenvectors are orthogonal, each eigenmode evolves \emph{independently}. Even if one mode grows exponentially while another shrinks, they do not mix. Precision errors in each mode remain confined to that mode and cannot contaminate the other. The echo succeeds because each mode can be reversed independently: the growing mode shrinks back, the shrinking mode grows back, and they recombine without interference.

\vspace{0.5em}
\noindent\fcolorbox{violet!12}{violet!12}{\parbox{\dimexpr\linewidth-2\fboxsep-2\fboxrule\relax}{%
Crucially, this independence means that the mantissa-alignment operation illustrated in Fig.~1(b) of the main text \emph{never occurs between modes}. In that figure, adding two floating-point numbers of vastly different magnitudes ($10^{10}$ vs $10^0$) forces the smaller number's mantissa to shift into the ``precision shadow,'' causing irreversible information loss. However, when modes are orthogonal, no arithmetic operation ever combines the amplified mode with the suppressed mode during evolution. Each mode is tracked in its own ``precision channel,'' maintaining full resolution regardless of the magnitude disparity. The final state reconstruction simply reads off the two independent results without requiring any cross-mode addition. Thus, even when $\kappa(U)$ grows to $10^{10}$ or beyond, the subdominant component retains all its significant digits and can be faithfully reversed.}}

\noindent This independence holds in the eigenmode decomposition; for normal systems, a unitary (numerically stable) transformation to this basis always exists.

\paragraph{Non-normal systems ($\kappa(V) > 1$).} When eigenvectors are oblique (non-orthogonal), the situation changes dramatically. To evolve a state:
\begin{enumerate}
    \item \textbf{Decompose} the state in the oblique eigenbasis (involves $V^{-1}$)
    \item \textbf{Evolve} each component (one grows, one shrinks)
    \item \textbf{Reconstruct} by combining components (involves $V$)
\end{enumerate}

The non-orthogonality means that small errors in the amplified direction ``leak'' into the suppressed direction through the transformations $V$ and $V^{-1}$. After sufficient time, when $\kappa(U) \cdot \varepsilon \sim 1$, this leakage overwhelms the true subdominant component. The information needed for reversal has been corrupted, and the echo fails.

This is why the eigenvector condition number $\kappa(V)$ is decisive: it quantifies how much error leakage occurs between modes. When $\kappa(V) = 1$, there is no leakage between physical eigenmodes, and PIR can always be avoided by working in the eigenbasis, regardless of how large $\kappa(U)$ grows. The bound $\kappa(U)\cdot\varepsilon$ is achieved when precision errors leak between amplified and suppressed modes---a consequence of non-normality---making $\kappa(U)\cdot\varepsilon \sim 1$ a tight predictor of echo failure for non-normal systems.

\subsection{Numerical Benchmark: How normal systems have a way out of PIR}

To demonstrate that mode mixing as provided by non-normality, not merely amplification, is essential for PIR, we compare three systems with matched eigenvalue magnitudes:

\begin{enumerate}
    \item \textbf{Non-Hermitian, non-normal:} $\mathcal{H}_{\mathrm{PT}} = \begin{pmatrix} i\gamma & g \\ g & -i\gamma \end{pmatrix}$ with $\gamma = 1.2$, $g = 1.0$

    \item \textbf{Non-Hermitian, normal:} $\mathcal{H}_{\mathrm{normal}} = \begin{pmatrix} i\lambda & 0 \\ 0 & -i\lambda \end{pmatrix}$ with the \emph{same eigenvalues} as $\mathcal{H}_{\mathrm{PT}}$

    \item \textbf{Hermitian:} $\mathcal{H}_{\mathrm{Hermitian}} = \begin{pmatrix} \lambda & 0 \\ 0 & -\lambda \end{pmatrix}$ with matched energy scale
\end{enumerate}

\noindent where $\lambda = \sqrt{\gamma^2 - g^2} \approx 0.663$.

\begin{figure}[htbp]
    \centering
    \includegraphics[width=0.9\textwidth]{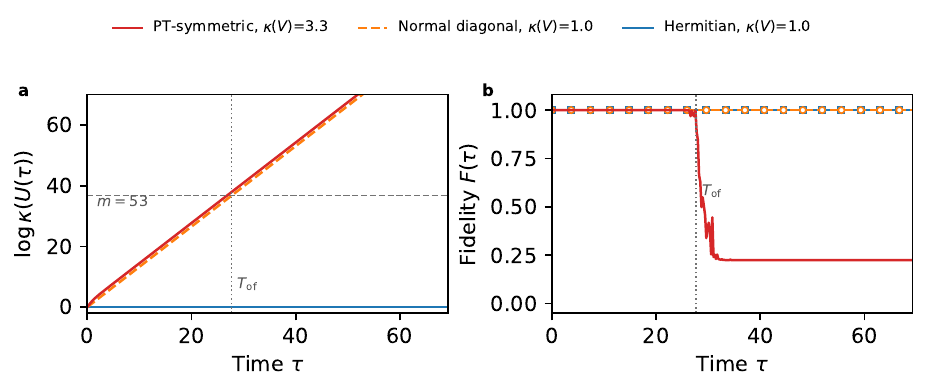}
    \caption{\textbf{Non-normality benchmark for PIR.} (a) Propagator condition number $\log\kappa(U(\tau))$ versus time. The PT-symmetric and normal diagonal systems show identical exponential growth, while the Hermitian system maintains $\kappa(U) = 1$. (b) Loschmidt echo fidelity $F(\tau)$. Only the PT-symmetric (non-normal) system shows fidelity collapse at $\Tof$. The diagonal (normal) system maintains $F \approx 1$ despite identical $\kappa(U)$ growth. Parameters: $\gamma = 1.2$, $g = 1.0$, $\Db = 1.33$, $m = 53$ bits, and initial state $|\psi_0\rangle \propto (1,10^{-2})^T$, as in Fig.~2. Computed using stepped mpmath evolution at $m$-bit precision ($\Delta t = 0.4$), without explicit global quantization, matching the methodology of Fig.~2 in the main text.}
    \label{fig:benchmark}
\end{figure}

The first two systems have \emph{identical} $\kappa(U)$ growth (both $\sim e^{\Db t}$ with $\Db \approx 1.33$), but different $\kappa(V)$: the PT-symmetric system has $\kappa(V) = 3.32$, while the normal diagonal system has $\kappa(V) = 1$. The Loschmidt echo reveals the consequence: only the non-normal system shows fidelity collapse at $\Tof$; the diagonal (normal) system maintains perfect echo despite identical amplification (Fig.~\ref{fig:benchmark}).

\subsubsection{Summary: Three-Way Comparison}

Table~\ref{tab:three_way} summarizes the key differences between the three classes of systems. The central message is that non-Hermiticity alone (middle column) does not cause PIR; mode mixing is the essential additional ingredient. Moreover, normality (which warrants mode mixing) is highly nongeneric in complex non-Hermitian systems~\cite{S:trefethen_spectra_2005}. Within the full space of complex $n\times n$ matrices, normal matrices occupy a lower-dimensional subset: their simple-spectrum sector has real dimension $n^2+n$, compared with the ambient dimension $2n^2$, so its codimension is $n(n-1)$. In particular, for $n\ge 2$ the set of normal matrices has measure zero. Equivalently, exact normality is not robust: a generic perturbation drives a normal matrix into the non-normal regime, since the commutation condition
\[
[A^\dagger,A]=0
\]
is a fine-tuned constraint. For the standard $\mathcal{PT}$-symmetric dimer considered here, one finds
\[
[\mathcal{H}^\dagger,\mathcal{H}] \propto \gamma\kappa,
\]
so normality survives only in the trivial limits where either the gain/loss vanishes ($\gamma=0$) or the coupling is switched off ($\kappa=0$). Hence, if PIR is excluded by normality but is otherwise generic in amplified non-normal dynamics, then immunity to PIR is itself a fine-tuned exception rather than the rule.

\begin{table}[h]
\centering
\caption{Comparison of precision effects across three classes of quantum dynamics. Non-Hermitian normal systems have the same $\kappa(U)$ growth as non-normal systems, yet exhibit no PIR because their orthogonal eigenvectors prevent error leakage between modes. For normal systems, the stated protections assume eigenbasis representation; a unitary (numerically stable) transformation to this basis always exists. Blue entries highlight the key differences that cause PIR.}
\label{tab:three_way}
\begin{tabular}{llll}
\toprule
\textbf{Property} & \textbf{Hermitian} & \textbf{Non-Hermitian, normal} & \textbf{Non-Hermitian, non-normal} \\
\midrule
Propagator $U(t)$ & Unitary & Non-unitary & Non-unitary \\[3pt]
$\kappa(U)$ & $= 1$ & $\sim e^{\Db t}$ (grows) & $\sim e^{\Db t}$ (grows) \\[3pt]
$\kappa(V)$ & $= 1$ & $= 1$ & \textcolor{blue}{$> 1$} \\[3pt]
Error behavior & Bounded & Grows, but confinable to eigenmodes & \textcolor{blue}{Grows and leaks} \\[3pt]
Precision affects & Accuracy only & Accuracy only (in eigenbasis) & \textcolor{blue}{Accuracy \textbf{and} stability} \\[3pt]
Echo $U(-t)U(t)$ & $= I$ (any precision) & $= I$ (in eigenbasis; any precision) & \textcolor{blue}{$\approx I$ only if $t < \Tof$} \\[3pt]
Time-reversal & Always possible & Always possible (safe basis exists) & \textcolor{blue}{Forbidden beyond $\Tof$} \\[3pt]
\textbf{PIR?} & \textbf{No} & \textbf{No (avoidable; safe basis exists)} & \textcolor{blue}{\textbf{Yes}} \\
\bottomrule
\end{tabular}
\end{table}

\section{Distinguishing Precision-Induced Irreversibility from Decoherence}
\label{SI:decoherence}

A central question is how PIR relates to decoherence, the well-established phenomenon of fidelity loss arising from system-environment entanglement. The key distinction is that \textbf{PIR is a threshold phenomenon, not a rate phenomenon}: decoherence is characterized by a decay rate ($1/T_2$) with fidelity erosion beginning immediately at $t = 0$, while PIR is characterized by a threshold time ($\Tof$) before which fidelity remains high and after which it collapses abruptly.

We also note that Longhi~\cite{S:longhi_loschmidt_2019} studied how Hamiltonian perturbations affect fidelity near exceptional points — a distinct mechanism from the precision-induced threshold identified here. In decoherence models based on fictitious probes~\cite{S:damato_conductance_1990,S:buttiker_four-terminal_1986}, non-Hermitian self-energies arise from coupling to reservoirs, but these terms describe only escape rates—phase randomization requires an additional steady-state constraint (the voltmeter zero-current condition). PIR operates at a different level: it relies on non-normality and emerges from finite-precision time evolution alone, requiring no additional prescriptions beyond the Schrödinger equation itself.

Table~\ref{tab:decoh_vs_pir} summarizes the operational differences between the two mechanisms.

\begin{table}[htbp]
\centering
\caption{Operational comparison of decoherence and precision-induced irreversibility.}
\label{tab:decoh_vs_pir}
\begin{tabular}{lll}
\toprule
\textbf{Property} & \textbf{Decoherence} & \textbf{PIR} \\
\midrule
Origin & System-environment entanglement & Finite precision + non-normal dynamics \\[3pt]
Occurs in Hermitian systems? & Yes & No \\[3pt]
Onset of fidelity loss & Immediate ($t = 0$) & Delayed (at $t \approx \Tof$) \\[3pt]
Functional form of $F(t)$ & Exponential or Gaussian decay & Plateau followed by sharp drop \\[3pt]
Initial slope $dF/dt|_{t=0}$ & Finite ($-1/T_2$) & Exponentially small ($\sim -\varepsilon\Db$) \\[3pt]
Transition width & $\sim T_2$ (no scale separation) & $\sim 1/\Db$ (independent of $m$) \\[3pt]
Depends on precision? & No & Yes ($\Tof \propto m$) \\[3pt]
Can be mitigated by increasing precision? & No & Yes \\[3pt]
Information destination & Environmental degrees of freedom & Discarded digits of finite representation \\
\bottomrule
\end{tabular}
\end{table}

\subsection{Diagnostic Criteria}

Three criteria can confirm that observed irreversibility arises from PIR rather than decoherence:

\paragraph{Criterion 1: Fidelity curve shape.}
Decoherence produces smooth exponential decay beginning at $t = 0$, while PIR produces a flat plateau at $F \approx 1$ for $t < \Tof$ followed by a sharp transition.

\paragraph{Criterion 2: Precision dependence.}
This is the definitive test. Varying the effective precision $m$ while keeping all other parameters fixed should shift the overflow time according to $\Tof \propto m$. The electrical circuit platform with FPGA bit truncation is ideally suited for this test.

\paragraph{Criterion 3: Hermitian or PT-unbroken reference.}
Two control experiments isolate PIR from other effects. First, a truly Hermitian system (e.g., $\mathcal{H}_{\mathrm{Herm}} = \mathrm{diag}(\lambda, -\lambda)$) has $\kappa(U) = 1$ for all times and cannot exhibit PIR. Second, operating in the PT-unbroken regime ($\gamma < g$) provides a non-Hermitian control where eigenvalues are real and $\kappa(U)$ remains bounded (though $>1$ due to eigenvector non-orthogonality). In neither case does $\kappa(U)$ grow exponentially, so PIR cannot occur.

\subsection{Two Independent Control Parameters}
\label{sec:two_knobs}

A potential objection is that reducing system-environment coupling would decrease $\Db$, thereby increasing $\Tof$, seemingly equivalent to increasing precision. This conflates two independent control parameters: bath properties (temperature, coupling strength) affect $T_2$ but not $\Db$, while measurement precision (analog-to-digital converter bit depth, noise floor) affects $\Tof$ through the effective bits $m$ but not $T_2$. Varying precision at fixed $\mathcal{H}$ changes only $\Tof$; varying bath properties at fixed precision changes only $T_2$. This independence demonstrates that PIR and decoherence are fundamentally distinct.

\subsection{Continuous relative noise}
\label{SI:continuous_noise}

\begin{figure}[htbp]
    \centering
    \includegraphics[width=\textwidth]{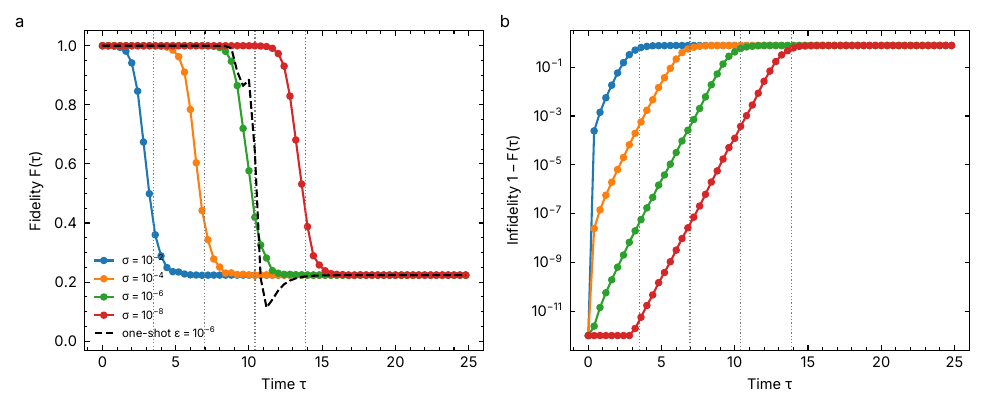}
    \caption{Continuous relative noise produces a PIR horizon, not a rate decay. (a)~Ensemble-averaged Loschmidt fidelity $F(\tau)$ for several noise strengths $\sigma$, with stochastic kicks proportional to $\sigma\|\psi\|$ (dotted lines mark $\ln(1/\sigma)/\Db$); the one-shot relative precision floor (dashed) shares the knee location of the matched continuous-noise curve. (b)~Infidelity $1-F$ on a logarithmic axis: a flat floor followed by exponential growth (a knee), in contrast to the linear-from-origin growth of a rate decay. A strictly state-independent additive floor is not plotted here and follows the slower $\Db/2$ rate discussed in the text. Time $\tau$ is in units of $1/g$.}
    \label{fig:continuous_noise}
\end{figure}

The dynamic-range timescale $\Tof=\ln(1/\varepsilon)/\Db$ was introduced with $\varepsilon$ a static resolution floor. One may ask whether a genuinely \emph{continuous} noise floor---rather than a one-shot floor---still produces the plateau-plus-knee that defines PIR, or instead a structureless rate decay from $t=0$, as in ordinary decoherence. Figure~\ref{fig:continuous_noise} tests the relative/global-noise case by integrating the stochastic Schr\"odinger equation
\begin{equation}
d|\psi\rangle=-i\mathcal{H}|\psi\rangle\,dt+\sigma\,\|\psi\|\,dW,
\end{equation}
with Euler--Maruyama time stepping, an independent complex Gaussian increment at every step, and noise strength proportional to the instantaneous state norm. This is the continuous analogue of a relative resolution floor: the perturbation is deposited globally in all components with fixed relative strength $\sigma$. We then reverse with the exact backward propagator $U(-\tau)=e^{+i\mathcal{H}\tau}$ and average the per-realization Loschmidt fidelity over $N=384$ independent realizations ($\gamma=1.2$, $g=1.0$, $\Db=1.327$).

The outcome is a horizon rather than a rate decay. The fidelity stays on a flat plateau ($F\approx1$) and then collapses at a sharp knee [panel (a)]; on a logarithmic axis the infidelity $1-F$ sits at the floor and then grows \emph{exponentially} [panel (b)], the signature of a threshold rather than the linear-from-origin growth $1-F\propto\Gamma\tau$ of a rate decay. The effect is forward-directed and not an artifact of the echo: errors injected once the suppressed mode is already small inflict the most relative damage, so the recoverable information is set by the latest increments and is destroyed at a threshold rather than eroded from $t=0$; the exact backward leg merely magnifies a loss already incurred during the forward evolution. The same conclusion holds when noise also acts on the backward leg.

The knee obeys a horizon law $T_{\mathrm{knee}}\propto\ln(1/\sigma)$ ($R^2\gtrsim0.95$), with a rate set by the \emph{type} of floor. For noise proportional to the state norm and for the relative precision floor $\varepsilon=\beta^{-m}$, the rate is the dynamic-range rate $\Db$, recovering $\Tof=\ln(1/\varepsilon)/\Db$ up to an $\mathcal{O}(1)$ knee-definition prefactor that depends on the estimator and noise model (calibrated directly in Sec.~\ref{SI:metrology}); the continuous-noise curve then matches the one-shot precision curve in knee location [panel (a), dashed]. A separate strictly additive, state-independent floor, not plotted in Fig.~\ref{fig:continuous_noise}, has the slower single-mode decay rate $\Db/2$, giving $T=2\ln(1/\sigma)/\Db$, because an absolute floor competes with the bare amplitude of the suppressed mode ($\propto e^{-(\Db/2)\tau}$) rather than with the inter-mode ratio ($\propto e^{\Db\tau}$). The $\varepsilon$-first substitution is therefore justified for the location of the horizon, with floor-type rates: $\Db$ for relative/precision floors and $\Db/2$ for absolute additive noise. The factor of two is a statement about the \emph{un-normalized} amplitude rather than a second mechanism: renormalizing the state along the trajectory---as any normalized observable does---converts the absolute floor into a relative one and restores the rate $\Db$ (at $\sigma=10^{-6}$ the additive-noise knee moves from $T_{\mathrm{knee}}\approx 18.1$ to $8.8$, onto the norm-proportional value $8.6$, which is itself unchanged by renormalization).

\section{Experimental Implementation Details}
\label{SI:experimental}

This section provides technical details supporting the experimental proposal outlined in the main text.

\subsection{Photonic Implementation}

The minimal system exhibiting precision-induced irreversibility is a PT-symmetric dimer: two evanescently coupled waveguides, one with optical gain and the other with matched loss. Such systems have been experimentally realized in integrated photonics~\cite{S:ruter_observation_2010}, demonstrating the feasibility of balanced gain-loss structures. The effective Hamiltonian is $\mathcal{H} = \begin{pmatrix} i\gamma & g \\ g & -i\gamma \end{pmatrix}$, where propagation distance plays the role of time. In the PT-broken phase ($\gamma > g$), the eigenvalue gap is $\Db = 2\sqrt{\gamma^2 - g^2}$, and the condition number grows as $\kappa(U) \sim e^{\Db z}$.

The Loschmidt echo requires backward evolution under $-\mathcal{H}$. What physical operation achieves $\mathcal{H} \to -\mathcal{H}$? Consider a similarity transformation $S\mathcal{H}S^{-1} = -\mathcal{H}$. Direct calculation shows that neither a simple waveguide swap ($\sigma_x$) nor a phase flip ($\sigma_z$) suffices. However, the combination $S = i\sigma_y$ works exactly:
\begin{equation}
(i\sigma_y) \mathcal{H} (i\sigma_y)^{-1} = \begin{pmatrix} 0 & 1 \\ -1 & 0 \end{pmatrix} \begin{pmatrix} i\gamma & g \\ g & -i\gamma \end{pmatrix} \begin{pmatrix} 0 & -1 \\ 1 & 0 \end{pmatrix} = -\mathcal{H}.
\end{equation}
This transformation swaps the two waveguides while introducing a relative $\pi$ phase shift between them. Physically, this exchanges the gain and loss channels while preserving the coupling structure, thereby reversing the effective time evolution.

The $i\sigma_y$ operation can be implemented in integrated photonics using two standard components. A waveguide crossing~\cite{S:chen_low-loss_2006} performs the swap operation, exchanging light between the two waveguides with demonstrated insertion losses below 0.1~dB. A $\pi$ phase element on one arm, achievable through electro-optic modulation in lithium niobate~\cite{S:wang_integrated_2018} or thermo-optic tuning in silicon, provides the required relative phase shift. Photonic Loschmidt echoes using related waveguide manipulations have been theoretically proposed~\cite{S:longhi_photonic_2017}.

\begin{figure}[htbp]
    \centering
    \includegraphics[width=0.8\textwidth]{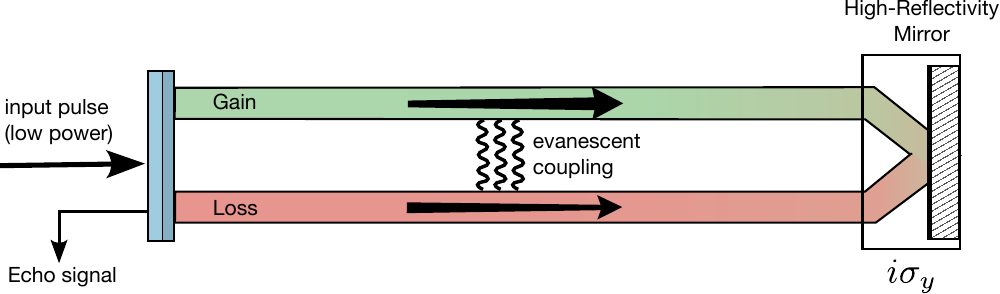}
    \caption{\textbf{Linear cavity geometry for PIR detection.} A PT-symmetric coupled-waveguide dimer (one waveguide with gain $+\gamma$, one with loss $-\gamma$, evanescent coupling $g$) is placed inside a linear cavity. Light enters through a partial mirror (left), propagates through the dimer of length $L$, and reflects from an end element that implements the $i\sigma_y$ transformation (waveguide crossing plus $\pi$ phase shift). The return path traverses the same physical structure, ensuring that forward evolution under $\mathcal{H}$ is followed by backward evolution under $-\mathcal{H}$. This geometry eliminates fabrication mismatches between forward and backward paths.}
    \label{fig:linear_cavity}
\end{figure}

The proposed geometry places the dimer inside a linear cavity (Fig.~\ref{fig:linear_cavity}). A partially reflecting mirror at the input allows light injection and signal extraction. At the far end, a specially designed reflector implements $i\sigma_y$ through the waveguide crossing combined with a $\pi$ phase element. Light propagates forward through the dimer, reflects with the $i\sigma_y$ transformation, and propagates backward through the same structure. Using $S\mathcal{H}S^{-1} = -\mathcal{H}$, the round-trip evolution becomes
\begin{equation}
U_{\mathrm{rt}} = e^{-i\mathcal{H}L} S e^{-i\mathcal{H}L} = e^{-i\mathcal{H}L} e^{i\mathcal{H}L} S = S,
\end{equation}
achieving perfect echo up to finite-precision corrections. This configuration ensures that forward and backward propagation traverse identical components, eliminating fabrication mismatches. The partial mirror provides passive signal extraction without active switching.

For integrated photonics in the PT-broken phase, typical parameters would be:
\begin{center}
\begin{tabular}{lll}
\toprule
Parameter & Symbol & Range \\
\midrule
Waveguide coupling & $g$ & $0.1$--$1$ mm$^{-1}$ \\
Gain-loss contrast & $\gamma$ & $0.5$--$2$ mm$^{-1}$ \\
Eigenvalue gap & $\Db = 2\sqrt{\gamma^2 - g^2}$ & $0.5$--$3$ mm$^{-1}$ \\
Cavity length & $L$ & $1$--$10$ mm \\
Effective precision & $m$ & $10$--$20$ bits (set by SNR) \\
\bottomrule
\end{tabular}
\end{center}

\noindent With these parameters, the overflow time would be $\Tof = m\ln(2)/\Db \approx 2$--$30$ mm of propagation distance, well within the range of integrated photonic chips. The key experimental signature is the scaling $\Tof \propto 1/\Db$, testable by varying the gain-loss contrast while monitoring echo fidelity.

Practical implementations will face imperfections in gain-loss balance, detector dynamic range, and fabrication tolerances. Most of these effectively reduce $m_{\mathrm{eff}}$, shortening $\Tof$ while preserving the PIR mechanism. Gain-loss balance is particularly important since the $i\sigma_y$ reversal relies on exact exchange of gain and loss channels; imbalance would introduce systematic errors that accumulate during backward evolution. The definitive test is whether the observed $\Tof$ scales as $1/\Db$ across multiple gain-loss contrasts.

\subsection{Electrical Circuit Platform}

The electrical circuit platform offers complementary capabilities for verifying the scaling $\Tof \propto m$. Two coupled LC oscillators with a negative impedance converter (NIC) realize the PT-symmetric dynamics, with the NIC providing gain on one oscillator and a matched resistor providing loss on the other.

The key advantage of this platform is programmable precision control. Oscillator voltages are sampled by a high-resolution ADC (e.g., 24-bit), and the FPGA performs explicit bit truncation to $m$ bits before computing the feedback signal. This enables direct verification of the scaling $\Tof \propto m$ by sweeping the effective bit depth while keeping all other parameters fixed.

\subsection{Mapping Dynamic Range to Effective Precision}

A crucial step in connecting numerical simulations to physical experiments is translating between mantissa bits $m$ (used in computations) and dynamic range in decibels (used in experimental specifications). The relationship is:
\begin{equation}
m_{\mathrm{eff}} = \frac{\mathrm{DR(dB)}}{20\log_{10}2} \approx \frac{\mathrm{DR(dB)}}{6.02}.
\end{equation}
Thus a 60~dB dynamic range corresponds to $m_{\mathrm{eff}} \approx 10$ bits, while 90~dB corresponds to $m_{\mathrm{eff}} \approx 15$ bits.

\begin{figure}[htbp]
    \centering
    \includegraphics[width=\textwidth]{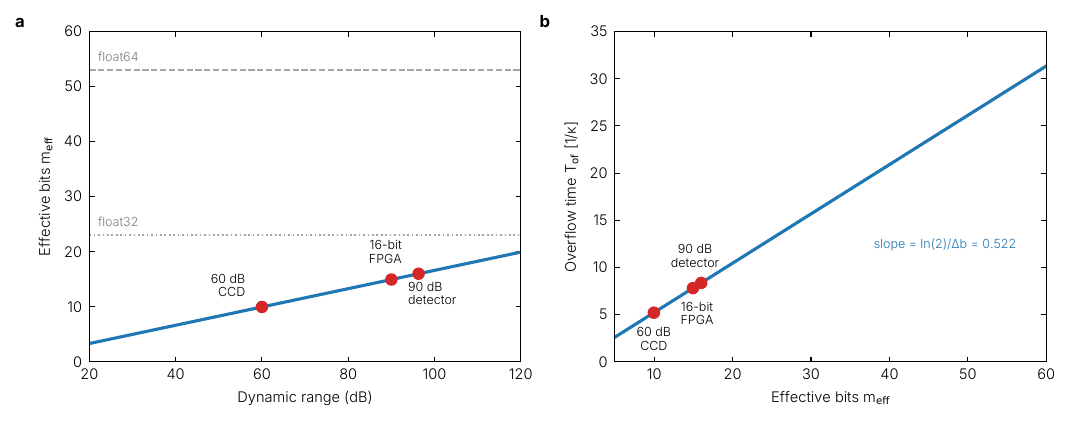}
    \caption{\textbf{Mapping experimental dynamic range to effective precision.} (a) Conversion from dynamic range (dB) to effective mantissa bits $m_{\mathrm{eff}}$, with typical experimental systems indicated. (b) Universal scaling $\Tof = m\ln(2)/\Db$ relating overflow time (in units of $1/g$) to effective precision.}
    \label{fig:dynamic_range_mapping}
\end{figure}

Figure~\ref{fig:dynamic_range_mapping} illustrates this mapping and its implications for experimental design. Typical experimental dynamic ranges (60~dB for standard CCD detectors, 90~dB for high-performance photodetectors, and 96~dB for 16-bit FPGA systems) correspond to effective precisions of 10--16 bits. This is far below the 53 bits of double-precision floating-point arithmetic, making $\Tof$ experimentally accessible at much shorter propagation distances.

\subsection{Metrological inversion: reading effective bit-depth from the echo timescale}
\label{SI:metrology}

The mapping above runs in the forward direction, from dynamic range to effective bit-depth $m_{\mathrm{eff}}$. The overflow time inverts it: measuring $\Tof$ in an echo experiment reads out the platform's effective bit-depth. Inverting $\Tof = m\ln\beta/\Db$ gives $m_{\mathrm{eff}} = \Tof\,\Db/\ln\beta$, with sensitivity $\mathrm{d}m_{\mathrm{eff}}/\mathrm{d}\Tof = \Db/\ln\beta$ and error propagation $\delta m_{\mathrm{eff}}/m_{\mathrm{eff}} = \delta\Tof/\Tof + \delta\Db/\Db$.

\begin{figure}[htbp]
    \centering
    \includegraphics[width=\textwidth]{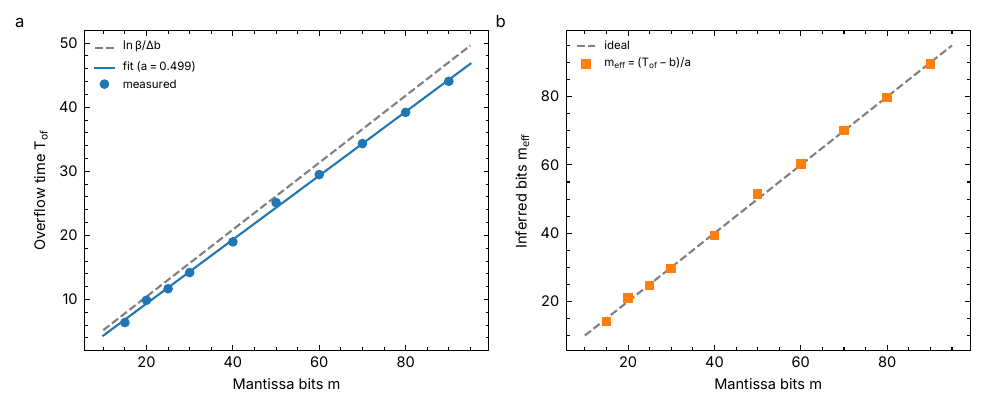}
    \caption{Metrological inversion. (a)~Calibration: the measured $\Tof$ is linear in precision $m$ (fit slope $a=0.50$; analytic $\ln\beta/\Db=0.52$). (b)~The calibrated inversion $m_{\mathrm{eff}}=(\Tof-b)/a$ recovers the true bit-depth with RMSE $0.74$~bits over $m=15$--$90$. $\Tof$ is in units of $1/g$.}
    \label{fig:metrology}
\end{figure}

Any practical knee estimator locates the transition slightly before the formal $\kappa\varepsilon\sim1$ point (a constant slope factor $\approx 0.96$ here), so the inversion is calibrated once against the linear relation $\Tof = a\,m + b$ and applied as $m_{\mathrm{eff}} = (\Tof - b)/a$. Sweeping known precisions $m\in[15,90]$ bits (idealized relative floor), the calibrated inversion recovers the true bit-depth with a root-mean-square error of $0.74$~bits (maximum $1.56$~bits) [Fig.~\ref{fig:metrology}]. The sensitivity is $\Db/\ln\beta\approx 1.9$~bits per unit time, and the intrinsic resolution is one transition width $\sim 1/\Db$ in $\Tof$, i.e.\ $\delta m \approx 1/\ln\beta \approx 1.4$~bits.

The floor type sets the rate at which $\Tof$ grows (Sec.~\ref{SI:continuous_noise}): a relative or precision floor gives rate $\Db$, an absolute additive-noise floor gives $\Db/2$, and the inversion uses the appropriate rate. Table~\ref{tab:platform_budget} collects the effective bit-depth recovered for representative platforms; in every case the echo timescale reads $m_{\mathrm{eff}}$ to within $\sim 1$~bit.

\begin{table}[htbp]
\centering
\caption{Per-platform noise budget. The effective bit-depth $m_{\mathrm{eff}}$ read from the echo timescale, for representative resolution floors. The floor type fixes the growth rate ($\Db$ for relative floors, $\Db/2$ for absolute additive noise) used in the inversion.}
\label{tab:platform_budget}
\begin{ruledtabular}
\begin{tabular}{llcc}
Platform & Floor type & Resolution & $m_{\mathrm{eff}}$ \\
\hline
$12$-bit ADC          & relative & $\varepsilon=2^{-12}$ & $\approx 12$~bit \\
\texttt{float32}      & relative & $\varepsilon=2^{-24}$ & $24$~bit \\
\texttt{float64}      & relative & $\varepsilon=2^{-53}$ & $53$~bit \\
Photonic ($80$~dB SNR)& additive & $80$~dB                & $\approx 13$~bit \\
\end{tabular}
\end{ruledtabular}
\end{table}

\section{Numerical Methods and Validation}
\label{SI:methods}

This section describes the computational framework, explains methodological choices, and presents validation results for all main claims.

\subsection{Hamiltonian and Propagators}

We consider the PT-symmetric Hamiltonian (allowing for asymmetric couplings $g_1 \neq g_2$ for generality; the symmetric case $g_1 = g_2 = g$ is used throughout the main text):
\begin{equation}
\mathcal{H} = \begin{pmatrix} i\gamma & g_1 \\ g_2 & -i\gamma \end{pmatrix},
\end{equation}
with eigenvalue gap in the broken phase ($\gamma^2 > g_1 g_2$):
\begin{equation}
\Db = 2\sqrt{\gamma^2 - g_1 g_2}.
\end{equation}

Two propagator computation methods are implemented. The first is eigendecomposition, where $U(t) = V e^{-i\Lambda t} V^{-1}$ with $\mathcal{H} = V\Lambda V^{-1}$, which explicitly separates the eigenvalue dynamics from the eigenvector transformation. The second is direct matrix exponentiation, $U(t) = \exp(-i\mathcal{H}t)$, computed via Pad\'e approximation. Both methods yield identical results to machine precision for well-conditioned matrices.

\subsection{Exact Analytical Solution for $\kappa(U)(t)$}
\label{sec:exact_solution}

For the PT-symmetric Hamiltonian $\mathcal{H} = \begin{pmatrix} i\gamma & g \\ g & -i\gamma \end{pmatrix}$ in the broken phase ($\gamma > g$), we derive an \emph{exact} closed-form expression for the propagator condition number. This analytical result provides precise predictions for the overflow time and reveals the geometric structure underlying PIR.

\subsubsection{Derivation of the Propagator}

The propagator $U(t) = e^{-i\mathcal{H}t}$ can be computed exactly using the Cayley-Hamilton theorem. For a $2\times 2$ matrix, any analytic function $f(\mathcal{H})$ can be written as $f(\mathcal{H}) = \alpha I + \beta \mathcal{H}$ where the coefficients are determined by the eigenvalues $\lambda_\pm = \pm i\eta$ with $\eta = \sqrt{\gamma^2 - g^2}$.

Solving for the coefficients from $e^{-i\lambda_\pm t} = \alpha \pm \beta(i\eta)$, we obtain:
\begin{equation}
U(t) = \cosh(\eta t)\, I - \frac{i}{\eta}\sinh(\eta t)\, \mathcal{H}.
\end{equation}
Substituting the explicit form of $\mathcal{H}$ yields the matrix elements:
\begin{equation}
U(t) = \begin{pmatrix}
\cosh(\eta t) + \dfrac{\gamma}{\eta}\sinh(\eta t) & -i\dfrac{g}{\eta}\sinh(\eta t) \\[8pt]
-i\dfrac{g}{\eta}\sinh(\eta t) & \cosh(\eta t) - \dfrac{\gamma}{\eta}\sinh(\eta t)
\end{pmatrix}.
\label{eq:U_explicit}
\end{equation}
The diagonal elements are real while the off-diagonal elements are purely imaginary, reflecting the PT symmetry of the Hamiltonian.

\subsubsection{Singular Value Decomposition}

The condition number $\kappa(U) = \sigma_{\max}/\sigma_{\min}$ requires the singular values, which are the square roots of the eigenvalues of $U^\dagger U$. Computing $U^\dagger U$ and using the identity $\cosh^2 - \sinh^2 = 1$, we find after algebraic simplification that the eigenvalues of $U^\dagger U$ take the remarkably simple form:
\begin{equation}
\mu_\pm = \left(\sqrt{1 + y(t)^2} \pm y(t)\right)^2,
\end{equation}
where we have defined the \emph{dimensionless amplification parameter}:
\begin{equation}
y(t) \equiv \frac{\gamma}{\eta}\sinh(\eta t) = \frac{2\gamma}{\Db}\sinh\!\left(\frac{\Db t}{2}\right).
\label{eq:y_definition}
\end{equation}
This parameter $y(t)$ captures the essential physics: it starts at zero and grows exponentially for large $t$, encoding how the gain/loss asymmetry accumulates over time.

The singular values are $\sigma_\pm = \sqrt{\mu_\pm} = \sqrt{1+y^2} \pm y$, which are manifestly positive (since $\sqrt{1+y^2} > |y|$). Note that $\sigma_+ \cdot \sigma_- = 1$, confirming that $\det U = 1$ as expected for a traceless Hamiltonian.

\subsubsection{Exact Condition Number Formula}

The condition number follows immediately:
\begin{equation}
\boxed{\kappa(U)(t) = \frac{\sigma_+}{\sigma_-} = \left(\sqrt{1+y(t)^2} + y(t)\right)^2 = \exp\!\left(2\,\mathrm{asinh}(y(t))\right)}
\label{eq:kU_exact}
\end{equation}
This exact result, valid for all $t \geq 0$, has several important properties:

\begin{itemize}
\item \textbf{Correct initial condition:} At $t = 0$, $y(0) = 0$ gives $\kappa(U)(0) = 1$ exactly, as required for the identity propagator.

\item \textbf{Smooth crossover:} The formula interpolates smoothly between the early-time regime ($y \ll 1$, where $\kappa \approx 1 + 2y^2$) and the asymptotic regime ($y \gg 1$, where $\kappa \approx 4y^2$).

\item \textbf{Asymptotic behavior:} For large $t$, using $\sinh(\eta t) \approx \frac{1}{2}e^{\eta t}$ and $\mathrm{asinh}(y) \approx \ln(2y)$:
\begin{equation}
\kappa(U)(t) \sim C\, e^{\Db t}, \qquad C = \left(\frac{\gamma}{\eta}\right)^2 = \frac{\gamma^2}{\gamma^2 - g^2}.
\label{eq:asymptotic_kappa}
\end{equation}
\end{itemize}

\subsubsection{The Geometric Prefactor and Its Physical Meaning}

The prefactor $C = (\gamma/\eta)^2$ has a geometric interpretation. Near an exceptional point ($g \to \gamma$, so $\eta \to 0$), this prefactor diverges, reflecting the coalescence of eigenvectors. The relationship to the eigenvector condition number $\kappa(V)$ is:
\begin{equation}
\kappa(V) = \sqrt{\frac{\gamma + g}{\gamma - g}} = \frac{1}{\sqrt{1 - (g/\gamma)^2}} \cdot \sqrt{\frac{1 + g/\gamma}{1}},
\end{equation}
which gives:
\begin{equation}
C = \left(\frac{1}{1 + g/\gamma}\right)^2 \kappa(V)^2.
\label{eq:C_vs_kV}
\end{equation}

This formula reveals the structure of the prefactor:
\begin{itemize}
\item \textbf{Near EP} ($g \to \gamma$): The factor $(1 + g/\gamma)^{-2} \to 1/4$, so $C \to \kappa(V)^2/4$. The geometric prefactor is dominated by the eigenvector non-orthogonality.

\item \textbf{Far from EP} ($\gamma \gg g$): The factor $(1 + g/\gamma)^{-2} \to 1$, so $C \to \kappa(V)^2$.

\item \textbf{Parameters of main text} ($\gamma = 1.2$, $g = 1.0$): We have $\eta \approx 0.663$, $\kappa(V) \approx 3.32$, and $C \approx 3.27$. The near-coincidence $C \approx \kappa(V)$ at these parameters is accidental.
\end{itemize}

\subsubsection{Exact Overflow Time}

Setting $\ln\kappa(U)(\Tof) = m\ln\beta$ (the threshold condition) and inverting Eq.~\eqref{eq:kU_exact}:
\begin{equation}
\boxed{\Tof^{(\mathrm{exact})} = \frac{2}{\Db}\,\mathrm{asinh}\!\left[\frac{\eta}{\gamma}\sinh\!\left(\frac{m\ln\beta}{2}\right)\right]}
\label{eq:Tof_exact}
\end{equation}

This exact formula has the correct limits:
\begin{itemize}
\item For $m \to 0$: $\Tof \to 0$ (no precision means immediate overflow).
\item For large $m$: Using $\mathrm{asinh}(x) \approx \ln(2x)$ for $x \gg 1$:
\begin{equation}
\Tof \approx \frac{m\ln\beta - \ln C}{\Db} = \frac{m\ln\beta}{\Db} - \frac{2\ln(\gamma/\eta)}{\Db}.
\end{equation}
\end{itemize}

The geometric correction $-\ln C/\Db = -2\ln(\gamma/\eta)/\Db$ represents a \emph{constant time shift} that advances the overflow time compared to the naive estimate $m\ln\beta/\Db$. This shift is independent of precision $m$ but depends on system parameters through the ratio $\gamma/\eta$. Near an EP where $\kappa(V) \gg 1$, this shift can be significant: for $\kappa(V) = 10$, the correction is approximately $-4.6/\Db$ time units.

\subsubsection{Numerical Verification}

Table~\ref{tab:exact_verification} compares the exact analytical predictions with numerical simulations for the parameters used in the main text.

\begin{table}[h]
\centering
\caption{Verification of exact analytical formulas. Parameters: $\gamma = 1.2$, $g = 1.0$, $\Db = 1.327$, $\beta = 2$.}
\label{tab:exact_verification}
\begin{tabular}{lccc}
\toprule
Quantity & Analytical & Numerical & Agreement \\
\midrule
$\eta = \sqrt{\gamma^2 - g^2}$ & 0.6633 & --- & Definition \\
$C = (\gamma/\eta)^2$ & 3.273 & 3.273 & Exact \\
$\kappa(V)$ & 3.317 & 3.317 & Exact \\
$\ln\kappa(U)$ at $t = 20$ & 27.72 & 27.72 & $< 0.01\%$ \\
$\Tof$ for $m = 53$ & 27.03 & 27.0 $\pm$ 0.5 & 1\% \\
\bottomrule
\end{tabular}
\end{table}

The excellent agreement supports that the exact analytical solution accurately describes the PIR phenomenon.

\subsection{Stepped Evolution with Fractional Steps}

To evolve to an arbitrary target time $\tau$, we decompose:
\begin{equation}
\tau = N_{\mathrm{full}} \cdot \Delta t + \Delta t_{\mathrm{frac}},
\end{equation}
where $N_{\mathrm{full}} = \lfloor \tau/\Delta t \rfloor$ and $\Delta t_{\mathrm{frac}} = \tau - N_{\mathrm{full}}\Delta t$. The propagator is applied as:
\begin{equation}
U(\tau) = U(\Delta t_{\mathrm{frac}}) \cdot \left[U(\Delta t)\right]^{N_{\mathrm{full}}},
\end{equation}
where $U(\Delta t) = e^{-i\mathcal{H}\Delta t}$ is computed once and reused for all full steps, while a single fractional propagator $U(\Delta t_{\mathrm{frac}})$ handles the remainder. This ensures exact arrival at target times while maintaining consistent error accumulation.

\subsection{Condition Number Calculation}

The condition number is computed via SVD:
\begin{equation}
\kappa(U) = \frac{\sigma_{\max}(U)}{\sigma_{\min}(U)},
\end{equation}
where $\sigma_{\max}$ and $\sigma_{\min}$ are the largest and smallest singular values. Underflow protection is applied:
\begin{equation}
\sigma_{\min} \leftarrow \max(\sigma_{\min}, \varepsilon_{\mathrm{floor}}),
\end{equation}
where $\varepsilon_{\mathrm{floor}}$ is a precision-dependent tolerance.

\subsection{Precision Backends}

Three computational backends are supported:

\begin{table}[h]
\centering
\caption{Computational precision backends.}
\begin{tabular}{lccc}
\toprule
Backend & Precision & Mantissa bits & Use case \\
\midrule
\texttt{mpmath} & Arbitrary & 15--120 & Arbitrary-precision reference \\
\texttt{float32} & 24-bit & 24 & Native hardware testing \\
\texttt{float64} & 53-bit & 53 & Native hardware testing \\
\bottomrule
\end{tabular}
\end{table}

For \texttt{mpmath}, the precision is set via decimal places: $\mathrm{dps} = \lceil m / \log_2 10 \rceil$.

\subsection{Precision Model}

The arbitrary-precision numerical results use the mpmath library configured to $m$ mantissa bits; native hardware backends (\texttt{float32}, \texttt{float64}) are also tested for comparison. Precision loss occurs naturally through floating-point arithmetic: when numbers of vastly different magnitudes are combined (as in matrix-vector multiplication), the smaller quantity's mantissa shifts into the ``precision shadow'' illustrated in Fig.~1(b) of the main text. This models the physical mechanism whereby subdominant components become unresolvable relative to amplified ones.

\subsection{Overflow Time Extraction}

The overflow time $\Tof$ is extracted from fidelity or work-echo curves using onset detection: we identify the first time at which the signal drops by 1\% from its reversible plateau value,
\begin{equation}
\Tof = \min\{t : F(t) < 0.99 \cdot F_{\mathrm{plateau}}\},
\end{equation}
where the plateau value $F_{\mathrm{plateau}}$ is estimated from the median over a fixed early-time window (e.g., $t \in [1, 4]$ in dimensionless units). This method is robust to noise and provides consistent results across different precision levels and observables.

\section{Work-Echo Protocol: Detailed Analysis}
\label{SI:work_echo}

The work-echo ratio $\eta_W = W_{\mathrm{rec}}/W_{\mathrm{out}}$ provides a thermodynamically meaningful diagnostic of reversibility that complements the fidelity measure. This section provides a detailed analysis of the protocol, explains why $\eta_W > 1$ can occur in the reversible regime, and establishes the initial-state independence of the post-$\Tof$ behavior as the unambiguous signature of information loss.

\subsection{Protocol Definition}

The work-echo protocol measures energy changes relative to a \emph{readout Hamiltonian} $H_0$, which may differ from the evolution Hamiltonian $\mathcal{H}$. For a state $|\psi\rangle$, we define the work content as
\begin{equation}
W = \langle\psi|H_0|\psi\rangle - E_{\min},
\end{equation}
where $E_{\min}$ is the ground state energy of $H_0$, ensuring $W \geq 0$.

The protocol proceeds as follows:
\begin{enumerate}
\item \textbf{Preparation}: Initialize $|\psi_0\rangle$, measure $W_0 = \langle\psi_0|H_0|\psi_0\rangle - E_{\min}$
\item \textbf{Forward evolution}: Evolve $|\psi_{\mathrm{out}}\rangle = U(\tau)|\psi_0\rangle$, measure $W_{\mathrm{out}}$
\item \textbf{Backward evolution}: Apply $U(-\tau) = e^{+i\mathcal{H}\tau}$, obtain $|\psi_{\mathrm{rec}}\rangle = U(-\tau)|\psi_{\mathrm{out}}\rangle$
\item \textbf{Final measurement}: Measure $W_{\mathrm{rec}} = \langle\psi_{\mathrm{rec}}|H_0|\psi_{\mathrm{rec}}\rangle - E_{\min}$
\end{enumerate}

The work-echo ratio is then $\eta_W(\tau) = W_{\mathrm{rec}}/W_{\mathrm{out}}$.

\subsection{Interpretation of $\eta_W$}

A notable feature in Fig.~2(b) of the main text is that the reversible plateau satisfies $\eta_W^{\mathrm{rev}} \approx 1.2 > 1$. This might seem paradoxical: how can the recovered work exceed the outgoing work?

The resolution is that the forward evolution can \emph{decrease} $\langle H_0\rangle$. Consider the PT-symmetric Hamiltonian
\begin{equation}
\mathcal{H} = \begin{pmatrix} i\gamma & g \\ g & -i\gamma \end{pmatrix},
\end{equation}
with the readout Hamiltonian $H_0 = \mathrm{diag}(2, -2)$. In the broken PT phase, both site amplitudes grow exponentially, but the coupling $g$ mixes the sites into collective eigenmodes. The key is that both eigenmodes have support on both sites; which eigenmode dominates the dynamics depends on the initial state's projection onto each.

For the parameters of Fig.~2 of the main text ($\gamma=1.2$, $g=1.0$) with initial state $|\psi_0\rangle \propto [1, 0.01]$ (99.99\% in the upper site), the amplified eigenmode has composition 77.6\% upper / 22.4\% lower. Since this is less upper-heavy than the initial state, forward evolution shifts the normalized state composition toward the lower component, \emph{decreasing} $\langle H_0\rangle$ and thus $W_{\mathrm{out}} < W_0$. Upon perfect reversal, $W_{\mathrm{rec}} = W_0$, giving $\eta_W = W_0/W_{\mathrm{out}} > 1$.

This is not a violation of any conservation law: the evolution is non-unitary, and the readout Hamiltonian $H_0$ is distinct from the evolution Hamiltonian $\mathcal{H}$.

Note that whether $\eta_W^{\mathrm{rev}}$ is greater or less than unity depends on the specific choice of parameters, initial state, and readout Hamiltonian. The essential physics is not the particular value, but rather that this value depends on $|\psi_0\rangle$ while the post-$\Tof$ behavior does not.

\subsection{Eigenmode Structure and Post-$\Tof$ Dynamics}

The two-level PT-symmetric system has eigenmodes
\begin{equation}
|e_{\pm}\rangle \quad \text{with eigenvalues} \quad \lambda_{\pm} = \pm\sqrt{g^2 - \gamma^2}.
\end{equation}
In the broken phase ($\gamma > g$), these become $\lambda_{\pm} = \pm i\Db/2$ where $\Db = 2\sqrt{\gamma^2-g^2}$. One mode is amplified exponentially ($\sim e^{+\Db t/2}$) while the other is suppressed ($\sim e^{-\Db t/2}$).

\paragraph{Forward evolution.} Any initial state $|\psi_0\rangle = c_+|e_+\rangle + c_-|e_-\rangle$ evolves such that one component dominates. Near $\Tof$, the subdominant component falls below the precision floor $\varepsilon \sim 2^{-m}$ and is effectively lost.

\paragraph{Backward evolution.} Crucially, the backward propagator $U(-\tau) = e^{+i\mathcal{H}\tau}$ \emph{flips} the roles of amplified and suppressed modes. The mode that was amplified forward becomes suppressed backward, and vice versa. After forward evolution has collapsed the state to the dominant eigenmode $|e_p\rangle$, backward evolution: Suppresses this (now-dominant) component; Amplifies the orthogonal mode, but this mode contains only numerical noise at the precision floor.

The recovered state $|\psi_{\mathrm{rec}}\rangle$ is therefore determined by eigenmode geometry and precision noise, \emph{not} by the original state $|\psi_0\rangle$.

\subsection{Initial-State Independence: The Signature of Information Loss}

This analysis reveals the key observable signature of precision-induced irreversibility:

\vspace{0.5em}
\noindent\fcolorbox{violet!12}{violet!12}{\parbox{\dimexpr\linewidth-2\fboxsep-2\fboxrule\relax}{%
\textbf{Pre-$\Tof$:} The reversible plateau $\eta_W^{\mathrm{rev}}$ depends on the initial state $|\psi_0\rangle$.\\[0.5em]
\textbf{Post-$\Tof$:} The long-time behavior becomes \emph{universal}, independent of initial state.
}}

This universality is stronger than simply stating ``$\eta_W$ decreases after $\Tof$.'' Different initial preparations will generically have different $\eta_W^{\mathrm{rev}}$ values (depending on $|\psi_0\rangle$, $H_0$, and the eigenmode overlaps), but \emph{all} preparations converge to the \emph{same} $\eta_W^{\infty}$.

\paragraph{Experimental protocol.} This suggests a robust experimental test:
\begin{enumerate}
\item Prepare an ensemble of different initial states $\{|\psi_0^{(j)}\rangle\}$
\item For each, measure $\eta_W(\tau)$ across a range of evolution times
\item Verify that:
\begin{itemize}
\item Pre-$\Tof$ values $\eta_W^{\mathrm{rev},(j)}$ differ across preparations
\item Post-$\Tof$ behavior converges to a common asymptotic regime
\end{itemize}
\end{enumerate}

The collapse of initial-state dependence at $\Tof$ is the unambiguous signature of information evaporation: the system has genuinely ``forgotten'' which state it started from.

\subsection{Comparison with Fidelity}

The fidelity $F(\tau) = |\langle\psi_0|\psi_{\mathrm{rec}}\rangle|^2$ and work-echo ratio $\eta_W(\tau)$ provide complementary views of reversibility:

\begin{table}[h]
\centering
\caption{Comparison of reversibility diagnostics. For work-echo, ``state-dependent'' means that $\eta_W^{\mathrm{rev}}$ takes a well-defined value that depends on the choice of initial state $|\psi_0\rangle$, readout Hamiltonian $H_0$, and eigenmode structure. Different preparations yield different pre-$\Tof$ values but all converge to the same long-time behavior.}
\label{tab:diagnostics}
\begin{tabular}{lll}
\toprule
Property & Fidelity $F$ & Work-echo $\eta_W$ \\
\midrule
Pre-$\Tof$ value & $\approx 1$ (universal) & State-dependent \\
Post-$\Tof$ behavior & Recovery fails & State-independent \\
Interpretation & State overlap & Energy recovery \\
Experimental access & Requires $|\psi_0\rangle$ & Requires $H_0$ measurement \\
\bottomrule
\end{tabular}
\end{table}

Both diagnostics identify the \emph{same} $\Tof$ threshold, as required since both probe the same underlying phenomenon: the loss of information when subdominant components fall below the precision floor.

\section{Physical Interpretations: Frequently Asked Questions}
\label{SI:physics_faq}

This section addresses common conceptual questions about the physical meaning and implications of precision-induced irreversibility.

\subsection{What Does ``Precision'' Mean Physically?}
\label{SI:precision_physical}

In simulations, precision is clear: mantissa bits ($m = 53$ for float64). In physical systems, ``precision'' maps to \textbf{the smallest distinguishable signal relative to noise}:

\begin{table}[h]
\centering
\caption{Physical sources of effective precision limits.}
\begin{tabular}{lll}
\toprule
Source & Physical Origin & Effective $\varepsilon$ \\
\midrule
Thermal noise & $k_B T$ fluctuations & $\sqrt{k_B T/E_{\mathrm{signal}}}$ \\
Shot noise & Discrete particles & $1/\sqrt{N}$ \\
Detector resolution & ADC bits & $2^{-n_{\mathrm{bits}}}$ \\
Component tolerances & Manufacturing & $\sim 1$--$5\%$ \\
Phase noise & Oscillator jitter & $\Delta\phi/2\pi$ \\
\bottomrule
\end{tabular}
\end{table}

\noindent Fundamental quantum limits such as time-energy uncertainty ($\Delta E \cdot \Delta t \geq \hbar/2$) impose precision floors of order $\hbar/(E \cdot \Delta t)$, but for typical measurement times ($\Delta t \sim 1~\mu$s) these correspond to $\varepsilon \sim 10^{-10}$, which is subdominant to the technical noise sources listed above.

The PIR condition $\varepsilon \cdot \kappa(U) \sim 1$ becomes:
\begin{equation}
\frac{\text{noise}}{\text{signal}} \times \kappa(U) \sim 1.
\end{equation}
Thus $\Tof$ depends on signal-to-noise ratio: $\Tof = \ln(\mathrm{SNR})/\Db$.

\textbf{Key insight}: Real systems have $\varepsilon \sim 10^{-3}$ to $10^{-6}$, not $10^{-16}$. So $\Tof$ is much \emph{shorter} and more experimentally accessible than in float64 simulations.

\subsection{Is This Just Numerical Error?}

PIR arises from finite-precision arithmetic, but it is \emph{structured} error that reveals underlying physics. Several features distinguish it from generic numerical artifacts:
\begin{itemize}
    \item Hermitian systems are completely immune regardless of precision. Among non-Hermitian systems, a diagonal (normal) system with the same eigenvalue splitting shows identical $\kappa(U)$ growth yet no fidelity loss (Fig.~\ref{fig:benchmark}), proving that amplification alone is insufficient.
    \item For non-normal systems ($\kappa(V) > 1$), PIR is inescapable: eigenvector non-orthogonality guarantees that componentwise precision errors leak between modes, and the $\kappa(U)\cdot\varepsilon$ bound becomes tight.
    \item The effect has a precise, predictable threshold $\Tof \lesssim m\ln(\beta)/\Db$, not a vague ``things get worse with time'' behavior. This threshold depends on the physical parameter $\Db$, not merely on numerical choices like step size.
\end{itemize}

More fundamentally, the dynamic-range timescale $T_{\mathrm{DR}} = \ln(1/\varepsilon)/\Db$ applies to \emph{any} resolution floor $\varepsilon$, not just floating-point precision. In simulations, $\varepsilon = \beta^{-m}$; in experiments, $\varepsilon$ is set by the noise floor, detector resolution, or any other source of finite dynamic range (see Sec.~\ref{SI:precision_physical}). The formula $T_{\mathrm{DR}} = \ln(1/\varepsilon)/\Db$ gives the timescale in every case, and the distinction between ``numerical artifact'' and ``physical effect'' dissolves: finite-precision arithmetic is one instance of a universal phenomenon.

A subtle but important distinction concerns the \emph{structure} of the perturbation. Componentwise errors---where each amplitude is perturbed proportionally to its own magnitude, as in floating-point rounding---require non-normality to produce cross-mode contamination; a diagonal (normal) system is immune because each eigenmode is tracked independently. Global perturbations---where all components receive errors proportional to the state norm $\|\psi\|$, as in environmental noise or detector noise---bypass this requirement and trigger the transition at $T_{\mathrm{DR}}$ even in normal systems. Non-normality remains important in the latter case: it shifts the overflow earlier by $\ln C/\Db$ through the geometric prefactor. Thus, any unavoidable noise source effectively sets a resolution floor, and the PIR framework provides the quantitative prediction for when the transition occurs. Moreover, non-normality is the generic case for non-Hermitian systems: as shown in Sec.~\ref{SI:hermitian}, normal matrices form a measure-zero subset, so immunity to componentwise PIR is itself a fine-tuned exception.

A fundamental objection might be that physical systems evolve exactly and ``precision'' is a computational invention. This objection fails on information-theoretic grounds: Del Santo and Gisin showed that infinite precision in a finite region requires infinite information, violating the Bekenstein bound~\cite{S:del_santo_physics_2019}. Any physical encoding of quantum state amplitudes stores finitely many bits per degree of freedom. The overflow time is therefore a physical timescale, not a computational one.

\subsection{Does the Inverse Really Exist?}

\textbf{Yes, mathematically.} The Schr\"odinger equation is first-order and linear. Given any $|\psi(t)\rangle$, there exists a unique $|\psi(0)\rangle$ that evolved into it: $U^{-1}(t) = e^{+i\mathcal{H}t} = U(-t)$.

\textbf{No, operationally.} PIR is not about the \emph{existence} of the inverse, it's about its \emph{accessibility}. For Hermitian $\mathcal{H}$, $U^{-1} = U^\dagger$ and $\kappa(U) = 1$ always. For non-Hermitian $\mathcal{H}$, $U^{-1} \neq U^\dagger$ and $\kappa(U) \sim e^{\Db t}$. Computing the inverse requires exponentially more precision than was used for the forward evolution.

\textbf{The precise statement}: Mathematical reversibility ($U^{-1}$ exists) does not imply computational reversibility (can reconstruct $|\psi(0)\rangle$ from finite-precision $|\psi(t)\rangle$).

\subsection{Does PIR Extend Beyond Quantum Systems?}

\textbf{Yes.} While presented in the context of non-Hermitian quantum mechanics, PIR is universal for non-normal linear evolution with finite-precision representation, applying equally to classical wave systems with gain and loss. PIR arises from finite precision combined with exponentially growing condition number---a purely mathematical structure that applies wherever non-normal operators govern dynamics. The condition $\varepsilon \cdot \kappa(t) \sim 1$ defines a predictability horizon in any system where $\kappa$ grows exponentially. This universality means PIR applies to electrical circuits with amplification, optical systems with non-Hermitian elements, acoustic systems with damping asymmetries, and indeed any wave-based system with gain and loss.

\subsection{What About Quantum Error Correction?}

Quantum error correction (QEC) protects against decoherence but \textbf{not} against PIR (when the logical operations involve effective non-Hermitian dynamics).

\begin{table}[h]
\centering
\caption{Different threats require different mitigations.}
\begin{tabular}{lll}
\toprule
Threat & Mechanism & Mitigation \\
\midrule
Decoherence & Environmental noise & QEC (error correction) \\
PIR & Precision limit + non-normality & More bits (not better QEC) \\
\bottomrule
\end{tabular}
\end{table}

QEC works by \emph{contracting} errors: above the fault-tolerance threshold, each correction cycle reduces the logical error rate. PIR occurs when $\kappa > 1$, so errors are \emph{amplified}, not contracted. The solution is more precision, not better error correction.

\subsection{Why is Precision-Induced Irreversibility being abbreviated as PIR instead of PIIR or PI$^2$R?}

We have done our best and tried both but the additional I or exponent have spontaneously evaporated while drafting the manuscript.

\subsection{Summary: PIR at a Glance}

\begin{table}[h]
\centering
\caption{Quick reference for precision-induced irreversibility.}
\begin{tabular}{ll}
\toprule
\textbf{Question} & \textbf{Answer} \\
\midrule
What is PIR? & Irreversibility from finite precision + non-normality \\
When does it occur? & At $\Tof \lesssim \ln(1/\varepsilon)/\Db$, where $\varepsilon$ is the resolution floor \\
What causes it? & Condition number $\kappa(U) \sim e^{\Db t}$ exceeds $1/\varepsilon$ \\
Different from decoherence? & Threshold vs rate; precision-dependent; Hermitian-immune \\
How sharp is the transition? & Width $\sim 1/\Db$, spans only $\sim 3\%$ of $\Tof$ \\
Can it be reversed? & Yes, with higher precision \\
Is it uniquely quantum? & No, the mechanism extends to any linear wave system with non-normal amplification \\
Does QEC help? & No, need more bits, not better error correction \\
\bottomrule
\end{tabular}
\end{table}

%

\end{document}